\documentclass[11pt,a4paper]{article}
\pdfoutput=1

\usepackage{jcappub}
\usepackage{graphicx}
\usepackage{epsfig}
\usepackage{subfigure}
\usepackage{comment}
\usepackage{slashed}
\usepackage{soul}
\usepackage{tabularx}
\usepackage{physics}
\usepackage[normalem]{ulem}
\usepackage{afterpage}
\usepackage{placeins}
\usepackage{booktabs}
\usepackage{float}

\newcommand{\mc}{\mathcal}

\usepackage{array}
\newcolumntype{C}[1]{>{\centering\let\newline\\\arraybackslash\hspace{0pt}}m{#1}}

\newcommand{\be}{\begin{equation}} 
\newcommand{\ee}{\end{equation}}
\newcommand{\bea}{\begin{equation}\begin{aligned}} 
\newcommand{\eea}{\end{aligned}\end{equation}}
\newcommand{\ber}{\begin{eqnarray}}
\newcommand{\ear}{\end{eqnarray}}

\def\lsim{\mathrel{\raise.3ex\hbox{$<$\kern-.75em\lower1ex\hbox{$\sim$}}}}
\def\gsim{\mathrel{\raise.3ex\hbox{$>$\kern-.75em\lower1ex\hbox{$\sim$}}}}

\newcommand{\Mpc}{{\rm Mpc}}

\newcommand{\GeV}{{\rm GeV}}

\newcommand{\ie}{{\it i.e.}}
\newcommand{\eg}{{\it e.g.}}

\newcommand{\td}{{\rm d}}
\newcommand{\Msun}{M_\odot}
\newcommand{\mpl}{M_{\rm P}}

\newcommand{\eps}{\epsilon}
\newcommand{\epsh}{\epsilon_{H}}
\newcommand{\epsv}{\epsilon_{V}}
\newcommand{\etah}{\eta_{H}}
\newcommand{\etav}{\eta_{V}}

\newcommand{\PR}{\mathcal{P}_{\mathcal{R}}}

\newcommand{\calH}{\mathcal{H}}

\definecolor{Red}{rgb}{1,0,0}
\definecolor{Blue}{rgb}{0,0,1}
\definecolor{Green}{rgb}{0,1,0}

\begin{document}

\title{Anatomy of single-field inflationary models for primordial black holes}

\author[a]{Alexandros Karam,}
\author[a]{Niko Koivunen,}
\author[a]{Eemeli Tomberg,}
\author[b]{Ville Vaskonen,}
\author[a]{and Hardi Veerm\"{a}e}

\affiliation[a]{NICPB, R\"avala 10, 10143 Tallinn, Estonia}  
\affiliation[b]{Institut de Fisica d'Altes Energies, Campus UAB, 08193 Bellaterra, Barcelona, Spain}

\emailAdd{alexandros.karam@kbfi.ee}
\emailAdd{niko.koivunen@kbfi.ee}
\emailAdd{eemeli.tomberg@kbfi.ee}
\emailAdd{vvaskonen@ifae.es}
\emailAdd{hardi.veermae@cern.ch}

\abstract{We construct an analytically solvable simplified model that captures the essential features for primordial black hole (PBH) production in most models of single-field inflation. The construction makes use of the Wands duality between the constant-roll (or slow-roll) and the preceding ultra-slow-roll phases and can be realized by a simple inflaton potential of two joined parabolas. Within this framework, it is possible to formulate explicit inflationary scenarios consistent with the CMB observations and copious production of PBHs of arbitrary mass. We quantify the variability of the shape of the peak in the curvature power spectrum in different inflationary scenarios and discuss its implications for probing PBHs with scalar-induced gravitational wave backgrounds. We find that the COBE/Firas $\mu$-distortion constraints exclude the production of PBHs heavier than $10^4 \Msun$ in single-field inflation.
}

\maketitle


\section{Introduction}
\label{sec:intro}

The existence of dark matter (DM) has been clearly established via its gravitational effect on cosmic structures, yet its nature and origin has eluded us for almost a century~\cite{Bertone:2016nfn}. The LIGO-Virgo collaboration's discovery of gravitational waves (GWs)~\cite{LIGOScientific:2016aoc, LIGOScientific:2018mvr, LIGOScientific:2020ibl, LIGOScientific:2021djp} originating from binary black hole (BH) mergers together with the absence of observational signatures of particle DM has reinvigorated interest on primordial black holes (PBHs)~\cite{Hawking:1971ei, Carr:1974nx} as a potential DM candidate~\cite{Bird:2016dcv,Clesse:2016vqa,Sasaki:2016jop}. However, numerous experiments have placed bounds on the PBH abundance (see \eg~\cite{Carr:2020gox} for a recent review) and PBHs are allowed to make up the entirety of DM only in the asteroid mass window $10^{17}\,{\rm g} \lesssim M_{\rm PBH} \lesssim 10^{21}\,{\rm g}$ \cite{Niikura:2017zjd, Katz:2018zrn, Montero-Camacho:2019jte}. Nevertheless, PBHs in the solar mass range could still contribute to the LIGO-Virgo merger events if solar mass PBHs make up about 0.1\% of the DM energy budget~\cite{Raidal:2017mfl,Ali-Haimoud:2017rtz,Raidal:2018bbj,Vaskonen:2019jpv,DeLuca:2020bjf, Hall:2020daa,Wong:2020yig,Hutsi:2020sol,DeLuca:2021wjr,Franciolini:2021tla,Franciolini:2021xbq,Franciolini:2022iaa}. PBHs heavier than $10^3 \Msun$ can provide seeds for supermassive BHs~\cite{Duechting:2004dk,Kawasaki:2012kn,Carr:2018rid} and play a role in generating cosmic structures~\cite{Carr:2018rid}. 

PBHs can form via various channels in the early Universe (see \eg~\cite{Khlopov:2008qy,Carr:2020gox} for reviews). 
The most commonly considered scenario involves the collapse of large density perturbations generated during inflation~\cite{
Carr:1975qj, Garcia-Bellido:1996mdl, Alabidi:2009bk, Kawasaki:2012wr, Clesse:2015wea, Garcia-Bellido:2016dkw, Carr:2017edp, Dimopoulos:2019wew, 
Ivanov:1994pa, Carr:1994ar, Bullock:1996at, Kawasaki:1997ju, Kawasaki:1998vx, Yokoyama:1998pt, Kohri:2007qn, Saito:2008em, Bugaev:2008bi, Drees:2011hb, Drees:2011yz, Kawasaki:2016pql, Garcia-Bellido:2017mdw, Domcke:2017fix, Ezquiaga:2017fvi, Kannike:2017bxn, Germani:2017bcs, Motohashi:2017kbs, Di:2017ndc, Ballesteros:2017fsr, Ballesteros:2020qam, Hertzberg:2017dkh, Cicoli:2018asa, Ozsoy:2018flq, Biagetti:2018pjj, Ezquiaga:2018gbw, Dalianis:2018frf, Gao:2018pvq, Rasanen:2018fom, Ballesteros:2018wlw, Carr:2018poi, Passaglia:2018ixg, Dalianis:2019asr, Atal:2019cdz, Drees:2019xpp, Kuhnel:2019xes, Bhaumik:2019tvl, Fu:2019ttf, Atal:2019erb, Arya:2019wck, Mahbub:2019uhl, Mishra:2019pzq, Ballesteros:2019hus, Ashoorioon:2019xqc, Fu:2019vqc, Fu:2020lob, Nanopoulos:2020nnh, Ozsoy:2020kat, Ragavendra:2020sop, Kefala:2020xsx, Ng:2021hll, Solbi:2021wbo, Inomata:2021uqj, Stamou:2021qdk, Wu:2021zta, Bastero-Gil:2021fac, Dalianis:2021iig, Solbi:2021rse, Zheng:2021vda, Cheng:2021lif, Teimoori:2021pte, Heydari:2021gea, Kawai:2021edk, Gangopadhyay:2021kmf, Rezazadeh:2021clf, Inomata:2021tpx, Heydari:2021qsr, Figueroa:2021zah, Wang:2021kbh, Iacconi:2021ltm, Sarkar:2021tsa, Cai:2021zsp, Cicoli:2022sih, Frolovsky:2022ewg}, but they may be produced also in cosmological phase transitions~\cite{Hawking:1982ga,Kodama:1982sf,Baker:2021nyl}, 
by the collapse of cosmic strings~\cite{Hawking:1987bn, Polnarev:1988dh, Garriga:1993gj, Caldwell:1995fu, MacGibbon:1997pu, Helfer:2018qgv, Jenkins:2020ctp}, 
false vacuum bubbles~\cite{Garriga:2015fdk,Deng:2017uwc,Deng:2020mds,Kusenko:2020pcg,Ashoorioon:2020hln,Maeso:2021xvl} 
or domain walls~\cite{Rubin:2001yw,Vachaspati:2017hjw,Ferrer:2018uiu,Ge:2019ihf,Deng:2016vzb}, 
or post-inflationary scalar field fragmentation~\cite{Cotner:2018vug,Cotner:2016cvr,Cotner:2019ykd}. 
These PBH formation scenarios could be probed indirectly via the associated GW background. In particular, large scalar perturbations during inflation induce a stochastic GW background via mode coupling at the second perturbative order~\cite{Matarrese:1993zf,Matarrese:1997ay,Nakamura:2004rm,Ananda:2006af,Baumann:2007zm} (for a recent review, see~\cite{Domenech:2021ztg}). This GW background has not been seen so far by LIGO-Virgo~\cite{Kapadia:2020pnr,Romero-Rodriguez:2021aws}, but it has been suggested that the recent signal from pulsar timing experiments at nHz frequencies~\cite{NANOGrav:2020bcs,Goncharov:2021oub,Chen:2021rqp,Antoniadis:2022pcn} could be related to GWs accompanying PBH formation from large inflationary scalar perturbations~\cite{Vaskonen:2020lbd,DeLuca:2020agl,Kohri:2020qqd,Domenech:2020ers}.

In this paper, we will focus on single-field inflationary scenarios for PBHs, which typically require potentials with specific features capable of inducing peaks in the curvature power spectrum. Over the past three decades, a large number of such models have been proposed~\cite{Ivanov:1994pa, Carr:1994ar, Bullock:1996at, Kawasaki:1997ju, Kawasaki:1998vx, Yokoyama:1998pt, Kohri:2007qn, Saito:2008em, Bugaev:2008bi, Drees:2011hb, Drees:2011yz, Kawasaki:2016pql, Garcia-Bellido:2017mdw, Domcke:2017fix, Ezquiaga:2017fvi, Kannike:2017bxn, Germani:2017bcs, Motohashi:2017kbs, Di:2017ndc, Ballesteros:2017fsr, Ballesteros:2020qam, Hertzberg:2017dkh, Cicoli:2018asa, Ozsoy:2018flq, Biagetti:2018pjj, Ezquiaga:2018gbw, Dalianis:2018frf, Gao:2018pvq, Rasanen:2018fom, Ballesteros:2018wlw, Carr:2018poi, Passaglia:2018ixg, Dalianis:2019asr, Atal:2019cdz, Drees:2019xpp, Kuhnel:2019xes, Bhaumik:2019tvl, Fu:2019ttf, Atal:2019erb, Arya:2019wck, Mahbub:2019uhl, Mishra:2019pzq, Ballesteros:2019hus, Ashoorioon:2019xqc, Fu:2019vqc, Fu:2020lob, Nanopoulos:2020nnh, Ozsoy:2020kat, Ragavendra:2020sop, Kefala:2020xsx, Ng:2021hll, Solbi:2021wbo, Inomata:2021uqj, Stamou:2021qdk, Wu:2021zta, Bastero-Gil:2021fac, Dalianis:2021iig, Solbi:2021rse, Zheng:2021vda, Cheng:2021lif, Teimoori:2021pte, Heydari:2021gea, Kawai:2021edk, Gangopadhyay:2021kmf, Rezazadeh:2021clf, Inomata:2021tpx, Heydari:2021qsr, Figueroa:2021zah, Wang:2021kbh, Iacconi:2021ltm, Sarkar:2021tsa, Cai:2021zsp, Cicoli:2022sih, Frolovsky:2022ewg}. We will briefly review them in the following.

The first toy model was developed in 1994 by Ivanov et al.~\cite{Ivanov:1994pa}, where they considered a potential with two power-law regions separated by a plateau. Similarly, an early toy model assumed a cliff-like feature where the slope becomes abruptly steeper~\cite{Bullock:1996at}. Although these models can produce large spikes in the power spectrum, they also predict a blue tilt at the CMB scales, which is now excluded. 

The first theoretically-motivated model was proposed in the context of supergravity~\cite{Kawasaki:1997ju} (and was further developed in~\cite{Kawasaki:1998vx, Kawasaki:2016pql}). In this scenario, a specific choice for the superpotential generates a polynomial inflaton potential, which, together with a complicated ``pre-inflation" phase, produces a power spectrum with a high amplitude and a shallow slope ($n_s < 1$) on small scales and a low amplitude and a nearly scale-invariant spectrum ($n_s \sim 1$) on large scales.

Inflation from a Coleman-Weinberg type potential $V \propto \phi^4 \ln(\phi^2/\mu^2)$ could produce two inflationary phases~\cite{Yokoyama:1998pt, Saito:2008em, Bugaev:2008bi}---chaotic inflation followed by crossing the minimum and a near-stop close to the origin which causes a second phase of inflation. A similar scenario is possible with the double well potential~\cite{Bugaev:2008bi} (see also~\cite{Fu:2020lob}). Unfortunately, the Coleman-Weinberg potential predicts a very large tensor-to-scalar ratio ($r > 0.3$) and a quite small scalar spectral index ($n_s < 0.94$). 

In the running mass model~\cite{Kohri:2007qn, Drees:2011hb, Drees:2011yz} (see also~\cite{Motohashi:2017kbs}), $V(\phi) = V_0 \pm m^2(\phi) \phi^2/2$, obtained from softly broken global supersymmetry, the power spectrum does not feature a peak, but instead it grows continuously on small scales. This feature can produce an overabundance of PBHs and, furthermore, result in eternal inflation unless a ``waterfall" field is introduced to force inflation to end.

More recently, a simple, theoretically well-motivated scenario involving a non-minimally coupled inflaton with a general quartic potential has been considered~\cite{Kannike:2017bxn, Ballesteros:2017fsr, Ballesteros:2020qam}. In this scenario, the Einstein frame possesses a quasi-inflection point (a small local maximum) for PBH production, and the non-minimal coupling leads to a plateau at large field values guaranteeing a small tensor-to-scalar ratio. However, non-minimal polynomial inflation with at most quartic terms will generally give a small scalar spectral index ($n_s < 0.95$) when it produces PBHs with lifetimes exceeding the age of the Universe~\cite{Kannike:2017bxn, Ballesteros:2017fsr, Ballesteros:2020qam}. This issue can be solved by quantum corrections, such as threshold effects in running couplings~\cite{Kannike:2017bxn} or effective higher-order operators~\cite{Ballesteros:2020qam}. A very similar scenario was initially considered in Ref.~\cite{Garcia-Bellido:2017mdw} (see also~\cite{Di:2017ndc, Hertzberg:2017dkh, Kuhnel:2019xes, Bhaumik:2019tvl, Ng:2021hll}), in which the Einstein frame potential of non-minimal quartic inflation was used, but without the field redefinition that arises when moving from the Jordan frame to the Einstein frame.

In the context of quasi-inflection point scenarios, it was shown that, around the small maximum, the slow-roll (SR) conditions fail, and the inflaton enters a temporary ultra-slow-roll (USR) phase during which it will slow down considerably~\cite{Kannike:2017bxn, Germani:2017bcs, Motohashi:2017kbs}. During this process, a sizeable peak in the spectrum can be generated. As a result, the usual SR approximation for the curvature power spectrum breaks down, and the Mukhanov-Sasaki equation must be solved numerically to obtain accurate results. In most cases, the potential parameters must be tuned to the precision of several significant digits as the field has to be balanced on top of a small local maximum for tens of $e$-folds before rolling over it.

Once it became widely understood that an asymptotically flat potential that features a quasi-inflection point at smaller field values can potentially produce PBHs and drive otherwise observationally successful inflation, a plethora of new models were suggested. Perhaps the most physically-motivated and economic of them is Critical Higgs Inflation~\cite{Ezquiaga:2017fvi} (see also~\cite{Ballesteros:2017fsr, Drees:2019xpp}). In this model, the non-minimally coupled Standard Model Higgs field is taken to be the inflaton. It has a simple quartic potential which gets logarithmic corrections once one considers the RG running of the quartic and non-minimal couplings. However, in~\cite{Rasanen:2018fom}, the model was studied carefully in both metric and Palatini formalisms and it was shown that one could obtain inflationary predictions in agreement with the experimental constraints only if $M_{\rm PBH} < 10^6 \ \rm g$.

Inflaton potentials with a quasi-inflection point can also arise in models of 
string inflation~\cite{Cicoli:2018asa, Ozsoy:2018flq, Cicoli:2022sih}, 
string-inspired axion monodromy~\cite{Ballesteros:2019hus}, 
superconformal attractors~\cite{Dalianis:2018frf, Dalianis:2019asr} (see also~\cite{Ragavendra:2020sop, Iacconi:2021ltm}), 
and no-scale supergravity~\cite{Gao:2018pvq, Nanopoulos:2020nnh, Stamou:2021qdk, Wu:2021zta}. Most of these models comply with the inflationary constraints and predict PBH DM in the asteroid mass window, $M_{\rm PBH} \sim 10^{-16}\ M_\odot - 10^{-13}\ M_\odot$. 

Large peaks in the power spectrum can also be induced by introducing a dump/dip\footnote{Note that introducing bumps/dips are mathematically equivalent to constructing potentials with small maxima (quasi-inflection points).} or a step to the potential. In the bumpy/dippy models~\cite{Atal:2019cdz, Atal:2019erb, Mishra:2019pzq, Zheng:2021vda, Rezazadeh:2021clf, Wang:2021kbh}, a small, often a Gaussian bump/dip multiplies a base potential. As bumps and dips can be placed anywhere in the potential, they can produce PBHs in a wide range of masses. Interestingly, a base potential with $n_s \geq 0.98$, excluded by CMB observations~\cite{Planck:2018jri}, becomes favoured when modified by the bump/dip. This is because the inflaton typically spends an extra $10-15$ $e$-folds during the power spectrum amplification near the bump/dip, reducing the length of the remaining SR period to at most 45-50 $e$-folds and thus results in a decrease in the value of $n_s$ as it is determined only by the latter. Similarly, one (or multiple) steep step(s) in the potential can lead to a strongly enhanced oscillating power spectrum~\cite{Kefala:2020xsx, Inomata:2021uqj, Dalianis:2021iig, Inomata:2021tpx, Cai:2021zsp}.

All models discussed above contain a feature in the potential. An alternative is to keep the potential simple and featureless but modify the gravitational component of the action beyond a simple non-minimal coupling. Examples of these models include $F(R)$ gravity~\cite{Frolovsky:2022ewg}, Dirac-Born-Infeld inflation~\cite{Ballesteros:2018wlw}, a non-minimal derivative coupling between the inflaton and gravity~\cite{Fu:2019ttf, Fu:2019vqc, Heydari:2021gea, Heydari:2021qsr}, Gauss-Bonnet inflation~\cite{Kawai:2021edk}, warm inflation~\cite{Arya:2019wck, Bastero-Gil:2021fac}, axion inflation with gauge fields~\cite{Ozsoy:2020kat}, $G$- or $k$-inflation~\cite{Solbi:2021wbo, Solbi:2021rse, Teimoori:2021pte}. Typically, in these models, the enhancement of the power spectrum arises due to its modified expression, which, apart from the potential, also depends on the extra non-minimal functions of the scalar field.

To sum up, inflationary models used for PBH production can be classified into two general categories:
\begin{enumerate}
    \item Models where the (Einstein frame) inflationary potential contains a feature. These, in turn, can be subdivided into three categories: 
    \begin{itemize}
        \item Small local maximum (which includes upward steps and quasi-inflection points)~\cite{Kannike:2017bxn, Ballesteros:2017fsr, Ballesteros:2020qam, Garcia-Bellido:2017mdw, Di:2017ndc, Hertzberg:2017dkh, Kuhnel:2019xes, Bhaumik:2019tvl, Ng:2021hll, Germani:2017bcs, Motohashi:2017kbs, Ezquiaga:2017fvi, Drees:2019xpp, Rasanen:2018fom, Cicoli:2018asa, Ozsoy:2018flq, Cicoli:2022sih, Ballesteros:2019hus, Dalianis:2018frf, Dalianis:2019asr, Ragavendra:2020sop, Iacconi:2021ltm, Gao:2018pvq, Nanopoulos:2020nnh, Stamou:2021qdk, Wu:2021zta, Atal:2019cdz, Atal:2019erb, Mishra:2019pzq, Zheng:2021vda, Rezazadeh:2021clf, Wang:2021kbh, Cai:2021zsp}
        \item Downward steps~\cite{Kefala:2020xsx, Inomata:2021uqj, Dalianis:2021iig, Inomata:2021tpx}
        \item Rolling through the global minimum~\cite{Yokoyama:1998pt, Saito:2008em, Bugaev:2008bi, Fu:2020lob}
        \item Potentials with multiple features stacked on top of each other, for instance, rapid oscillations~\cite{Cai:2019bmk} (see also~\cite{Tasinato:2020vdk}).
    \end{itemize}
    \item Modified gravity (beyond non-minimally coupled inflaton) with features in terms other than the inflaton potential or the non-minimal coupling~\cite{Frolovsky:2022ewg, Ballesteros:2018wlw, Fu:2019ttf, Fu:2019vqc, Heydari:2021gea, Heydari:2021qsr, Kawai:2021edk, Arya:2019wck, Ashoorioon:2019xqc, Bastero-Gil:2021fac, Ozsoy:2020kat, Solbi:2021wbo, Solbi:2021rse, Teimoori:2021pte}.
\end{enumerate}

The general behaviour of the power spectrum during the rapid transition from SR inflation to non-attractor inflation is of great importance for the resulting PBH phenomenology. Consequently, it is desirable to have analytic models for said behaviour. This is possible if one considers toy models in which the evolution of SR parameters is postulated instead of derived from solving the field equations. For example, in~\cite{Byrnes:2018txb} it was shown that in single-field inflation, independently of the shape of the inflaton potential, the steepest possible growth of the primordial power spectrum is $n_s - 1 = 4$, or equivalently, $k^4$, with $k$ being the comoving scale. A further refined analysis~\cite{Carrilho:2019oqg}, however, showed that a steeper $k^5 \left( \log k \right)^2$ growth can be possible (see also~\cite{Cai:2019bmk, Liu:2020oqe, Ozsoy:2019lyy, Ballesteros:2020sre, Ragavendra:2020sop, Tasinato:2020vdk, Ng:2021hll, Ozsoy:2021pws}). However, growth steeper than $k^4$ will not have any measurable impact on the PBH mass function~\cite{Cole:2022xqc} and, as we will show, requires an initial phase of inflation with a blue-tilted spectrum.

In this paper, we consider the general class of single-field inflation models in which an Einstein frame can be defined. We outline the essential qualitative features of these scenarios and propose a novel, analytically solvable model capable of capturing these features. Our analytic method uses the Wands duality~\cite{Wands:1998yp} between the USR and the subsequent constant-roll (CR) (or SR) phase, which is present in most single-field models for PBHs. Thus, it improves upon existing analytic approaches built around specific ans\"atze for the second SR parameter~\cite{Byrnes:2018txb,Carrilho:2019oqg,Cole:2022xqc}. Moreover, we show that our analytic model corresponds to a continuous potential constructed by gluing together two parabolas and can thus be seen as a limiting case of realistic inflationary scenarios. It can also be thought of as a generalization of the Starobinsky model~\cite{Starobinsky:1992ts}, in which the potential consists of two linear segments with different slopes. Although we will not consider models with multiple features, the proposed analytic approach can be straightforwardly generalized to include such scenarios.

Having approximate analytic tools at hand is beneficial not only for theoretical studies but also for inflationary model building and PBH phenomenology. Using various shapes for the power spectra, we analyse the potential of future GW experiments for probing the GW background induced by the large scalar perturbations in single-field inflation, showing that the windows where the PBHs could constitute all DM and where they could contribute to the LIGO-Virgo GW events, can be probed with the future GW interferometers and pulsar timing arrays, respectively~\cite{Saito:2008jc,Assadullahi:2009jc,Bugaev:2010bb,Alabidi:2012ex,Inomata:2016rbd,Orlofsky:2016vbd,Espinosa:2018eve,Inomata:2018epa,Byrnes:2018txb,Cai:2018dig,Bartolo:2018rku,Wang:2019kaf,Yuan:2019udt,Chen:2019xse,Lewicki:2021xku}. These analytical tools can also be applied to improve limits on heavy PBH scenarios, \eg, by using COBE/Firas $\mu$-distortion constraints~\cite{Fixsen:1996nj,Chluba:2012we,Chluba:2013dna}.

The paper is structured as follows: Section~\ref{sec:theory} outlines the general mathematical framework of inflation, cosmological perturbations and the general phases of single-field inflationary models for PBHs. In section~\ref{sec:insta}, we introduce the instantaneous SR to USR approximation, derive the corresponding analytic curvature power spectrum, and construct the corresponding inflationary potential. Non-instantaneous SR to USR transitions and the application of our analytic approximations to the study of spectra for general potentials are discussed in section~\ref{sec:spectrum_approximation}. Section~\ref{sec:pheno} deals with the resulting PBH and GW phenomenology. We conclude in section~\ref{sec:concl}. We use natural units $c=\hbar=1$.

\section{General considerations}
\label{sec:theory}

Inflationary PBH production within single-field inflation typically requires for inflation to proceed in two or more phases of SR or constant-roll (CR), separated by a brief exit from SR and a transient USR phase. The latter is responsible for the super-horizon enhancement of the scalar modes that exited the cosmological horizon around this transient phase~\cite{Leach:2000yw,Leach:2001zf} and will ultimately determine the position and height of the peak in the power spectrum responsible for PBH creation. The slope of the enhanced peak at the end of the USR phase generally mimics the spectrum created in the subsequent CR phase due to the Wands duality~\cite{Wands:1998yp}. This makes it possible to infer the qualitative features of the power spectrum, \eg, the height and the position of the generated peak, from background evolution alone, similarly to SR inflation. However, an accurate estimate of the shape of the peak requires a numerical approach to the evolution of individual modes, \ie, via applying the Mukhanov-Sasaki (MS) equation.

\subsection{Background evolution in single-field inflation}

In the Einstein frame, single-field inflation is described by the following action:
\be \label{eq:S_standard}
	\mc{S}=\int{\td x^4 \sqrt{-g} \left( \frac{\mpl^2}{2}R-\frac{1}{2}\partial^{\mu} \phi \partial_{\mu} \phi-V(\phi) \right)}\,,
\ee
where $R$ is the Ricci scalar, $V(\phi)$ is the inflaton potential and $\mpl$ the reduced Planck mass. We consider a flat FLRW background geometry $\td s^2= -\td t^2+a^2 \td x_i^2$, where $a$ is the scale factor.  The evolution of the scale factor and the inflaton are governed by the Friedmann equation $3\mpl^2H^2= \dot{\phi}^2/2+V(\phi)$, and the Klein-Gordon equation $\ddot{\phi} + 3 H \dot{\phi} + V' = 0$, respectively. The overdot indicates differentiation with respect to the cosmic time $t$, a prime denotes differentiation with respect to the scalar field $\phi$ and $H\equiv \dot{a}/a$ is the Hubble parameter. This approach includes all models in which the Einstein frame exists, \eg, non-minimally coupled inflatons or non-canonical inflaton kinetic terms. Theories with higher-order Einstein frame kinetic terms, such as $(\partial \phi)^4$, may be approximated by this class, if these higher-order terms are subdominant even when SR is violated, \eg, as in Palatini $R^2$ gravity~\cite{Enckell:2018hmo, Antoniadis:2018ywb, Tenkanen:2020cvw, Karam:2021sno}.  Permitting disformal transformations~\cite{Bekenstein:1992pj, Minamitsuji:2014waa, Tsujikawa:2014uza, Domenech:2015hka}, this also includes some Horndeski theories~\cite{Sato:2017qau, Gialamas:2020vto}.

In the following, we will measure time in $e$-folds, $N \equiv \ln a$. The field equations can be then be recast as
\bea\label{eq:eom_inf_N}
	\partial_N y + \left(3-\frac{y^2}{2}\right)\left(y + \mpl \frac{V'}{V}\right) = 0,	\qquad 
	y \equiv \frac{\partial_N \phi}{\mpl} \,,
\eea
and the Hubble parameter can be expressed as a function of $\phi$ and $y$ via $\mpl^2H^2 = V(\phi)(3-y^2/2)^{-1}$. The SR approximation is obtained by ignoring the $\partial_{N}y$ term in Eq.~\eqref{eq:eom_inf_N} and the field equations are given by setting the last bracket in \eqref{eq:eom_inf_N} to zero, \ie , $y + \mpl\partial_{\phi}\ln V = 0$. SR and deviations from it are conveniently described in terms of the Hubble SR parameters
\be\label{HRSPs}
    \epsh \equiv \frac{1}{2}y^2 = -\frac{\dot H}{H^2} \,, \qquad
    \etah = -\partial_{N}\ln y + \epsh = -\frac{\ddot H}{2H \dot H} \, .
\ee  
Inflation occurs as long as $\epsh < 1$, while the SR approximation holds as long as both SR parameters are small ($\epsh, \etah \ll 1$). For future reference we will also define the potential SR parameters
\be\label{PRSPs}
	\epsv \equiv \frac{\mpl^2}{2} \left( \frac{V'}{V} \right)^2 \,, \qquad
	\eta_{\rm V} \equiv \mpl^2 \, \frac{V''}{V} \, \, ,
\ee
which satisfy $\epsv \approx \eps_{H} \,, \etav \approx \etah + \epsh\,$ at the leading order in SR.

\subsection{Evolution of the curvature perturbations}
\label{sec:perturbation_evolution}

The evolution of the scalar curvature perturbation $\mc{R}_c$ is conveniently studied using the so-called Mukhanov field $u \equiv z \mc{R}_c$, where 
\be
    z \equiv a \dot{\phi}/H =  a y \mpl \, .
\ee
The Fourier modes of $u$ evolve according to the MS equation~\cite{Mukhanov:1985rz,Sasaki:1986hm,Mukhanov:1990me},
\be\label{eq:MS}
	u_{k}'' + \left(  k^2 -\frac{z''}{z} \right) u_{k}=0\,,
\ee
where the prime denotes differentiation with respect to conformal time\footnote{Not to be confused with $V'$ in which case the prime corresponds to a derivative with respect to the field.}, $\td \tau = \td t / a$. The second term in the parenthesis of the MS equation \eqref{eq:MS} is recast in terms of the SR parameters as\footnote{In terms of the higher-order Hubble SR parameters, $\eps_{i+1} \equiv \partial_{N}\ln \eps_{i}$ with $\eps_1 \equiv \epsh$, the effective mass can be recast as $z''/z = \mathcal{H}^2 \left( 2- \epsilon_1 + 3\eps_2/2 - \eps_1 \eps_2/2 + \eps_2^2/4 + \eps_2\eps_3/2 \right)$. In Eq.~\eqref{eq:zpp}, $\eps_2$ is replaced with $\etah$, while $\eps_3$ is eliminated using \eqref{eq:eom_inf_N} in favor of the potential SR parameter $\etav$.}
\bea\label{eq:zpp}
	\frac{z''}{z}
	&= \calH^2\left[ 2-\left(3-\epsh\right) \etav+\epsh \left(5+2\epsh-4\etah\right) \right] \, ,
\eea
where $\calH \equiv a'/a= aH$. We stress that this expression is exact to all orders in the SR parameters. In the SR phase, the SR parameters are small, implying $z''/z \approx 2\calH^2$.  The limits $k^2 \gg z''/z$ and $k^2 \ll z''/z$ correspond to the sub-horizon and super-horizon evolution of the mode, respectively. Both $z$ and $z''/z$ are monotonously growing functions during SR, so these limits occur at $N\to -\infty$ and $N\to +\infty$, respectively. However, if the inflationary universe contains a SR-violating phase such as USR, which is generally needed for producing a non-negligible PBH abundance, then $z$ may temporarily decrease, affecting the modes that exit the horizon around that time. Nevertheless, both $z$ and $z''/z$ will still be monotonously growing functions of time in the asymptotic past and future.

In the sub-horizon regime, when $k^2 \gg z''/z$,  the effect of the curvature of spacetime is negligible and thus the mode $u_k$ behaves as a free field on flat spacetime. In the Bunch-Davies vacuum, the normalized solution asymptotes to
\be\label{eq:BD_cond}
	u_{k}= \frac{1}{\sqrt{2 k}}e^{-ik \tau}\,
\ee
in the limit $\tau \to -\infty$. This serves as the initial condition for the MS equation.
\begin{figure}[H]
    \centering
    \includegraphics[width=0.97\textwidth]{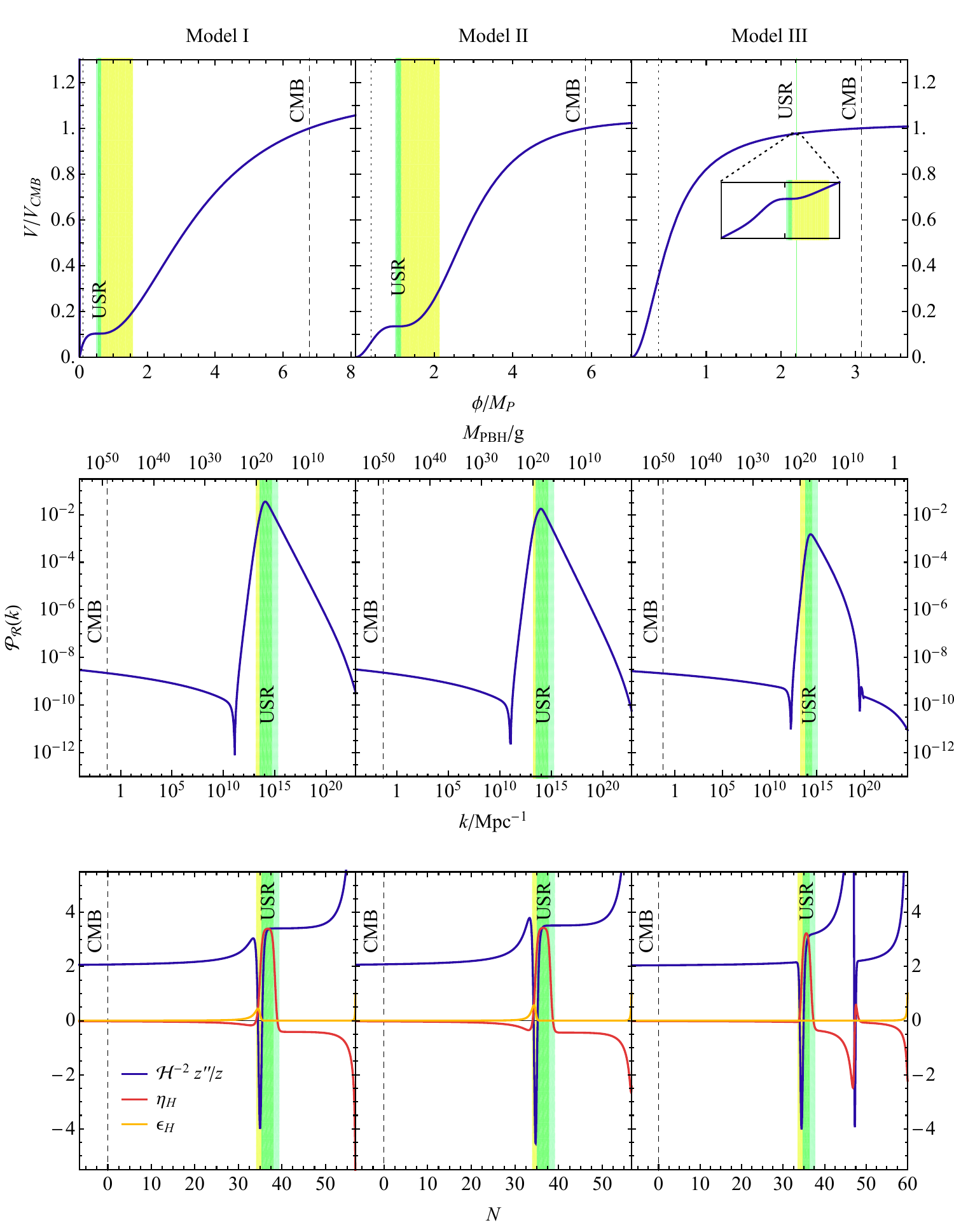}
    \caption{Examples of three inflationary models from the literature producing PBHs in the asteroid mass window~\cite{Kannike:2017bxn, Ballesteros:2017fsr, Ballesteros:2020qam, Dalianis:2018frf, Mishra:2019pzq}. The models and their parameters are detailed in appendix~\ref{app:examples}. The vertical bands correspond to different non-SR epochs: the USR epoch (green), the transition from SR to USR (yellow) and the transition out of USR to the subsequent CR phase (light green). For the power spectrum, they highlight the modes that exited during the corresponding epoch. The dashed vertical line corresponds to the CMB scale. The dotted vertical line in the top panel corresponds to the end of inflation; in the other panels, the end of inflation coincides with the right edge of the plot. In the bottom panel, $N$ denotes the number of e-folds of expansion starting from the CMB scale. All three potentials exhibit a local maximum around the USR scales.}
    \label{fig:example_models}
\end{figure}
In the super-horizon regime, that is when $k^2 \ll z''/z$, the general solution $u_k = A_{1,k} u^{(1)}_0 + A_{2,k} u^{(2)}_0$ comprises a mode that grows during SR, here $u_{0}^{(1)}$, and a decaying mode $u_{0}^{(2)}$, given by
\be\label{eq:SH_modes}
	u^{(1)}_0 \propto z, \qquad
	u^{(2)}_0 \propto z \int^{0}_{\tau} \frac{\td \tau'}{z(\tau')^2}.
\ee
The former will eventually dominate. Consequently, the amplitude of the curvature perturbations is frozen, that is,  $|\mc{R}_{c\, k}| =\left\lvert u_k/z\right\rvert ={\rm const}$. As alluded to above, if SR is momentarily violated in such a way that $z$ starts to decrease, then the growing and decreasing $u_k$ modes can be mixed. Therefore, the second mode may come to dominate temporarily, potentially leading to super-horizon evolution of $\mc{R}_c$; however, eventually $z$ starts to grow again and $\mc{R}_c$ freezes. This makes it possible to relate the observable perturbations at horizon re-entry with the perturbations produced during inflation while being ignorant about the details of the reheating phase.

The primordial power spectrum is evaluated after horizon crossing
\be\label{eq:PR_from_u}
	\PR
	=  \left. \frac{k^3}{2 \pi^2}\frac{|u_{k}|^2}{z^2} \right|_{\calH \gg k} 
	\stackrel{\rm SR}{\approx} \left. \frac{H^2}{8\pi^2 \mpl^2 \epsh} \right|_{k=\calH} \, ,
\ee
where the second expressions shows the result at leading order in the SR approximation, in which case $|u_{k}|^2 \to (\calH/k)^2/(2k)$ and $y$ are evaluated at horizon crossing, \ie , when $k=\calH$. During SR, the Hubble SR parameters~\eqref{HRSPs} can be replaced by the potential ones~\eqref{PRSPs}, and the scalar spectral index $n_s$ and the tensor-to-scalar ratio $r$ are given by the well known expressions
\be \label{eq:SR_ns_and_r}
    n_s 
    \approx 1 + 2\etav - 6\epsv \, , \qquad 
    r 
    \approx 16\epsv \, ,
\ee
at the leading order in SR. The curvature power spectrum is relatively unconstrained, except at scales relevant for the CMB. The latest observational constraints~\cite{Planck:2018jri,BICEP:2021xfz} imply
\bea \label{eq:CMB_observables}
    A_s &= (2.10 \pm 0.03)\times 10^{-9} , \quad
    n_s = 0.9649 \pm 0.0042, \quad
    r < 0.036 \qquad 
\eea
at the scale $k_{\rm pivot} = 0.05\ \Mpc^{-1}$ and at $68 \%$ CL for $A_s$ and $n_s$ and $95 \%$ for $r$. When running of $n_s$ is permitted, the constraints are slightly relaxed to $n_s = 0.9625 \pm 0.0048$. 
Throughout this paper, we will fix $A_s \equiv \PR(k_{\rm pivot}) = 2.1\times 10^{-9}$, when the power spectrum predictions involve the CMB scales.

\subsection{Phases of single-field inflation for PBHs}
\label{sec:phases}

In the models under consideration, we identify the following phases of inflation:
\begin{description}
    \item[SR] We will consider models in which inflation begins with a \emph{SR phase}.\footnote{Although other types of initial inflationary phases are permissible and would not significantly affect our results, we will label this initial phase as SR for definiteness.} At almost all times during this phase, the SR parameters remain small $\epsh,|\etah| \ll 1$ and $\etah$ slightly negative. For definiteness, we define the initial SR phase via the conditions
    \be\label{eq:SR_cond}
        |\epsh|, \, |\etah|, \, |\eps_3| < 1\, ,
    \ee
    where $\epsilon_3\equiv \partial_N \ln (\epsh - \etah)$ denotes the third SR parameter, which takes large values in particular when $\eta_H$ changes rapidly. With these conventions, we find that the initial SR phase typically ends by violating the $|\epsilon_3| < 1$ condition in PBH models.
    
    Note the enhancement of modes that exit at the end of the SR phase in Fig.~\ref{fig:example_models}. This is due to super-horizon effects induced during the subsequent phases~\cite{Leach:2001zf}, that is, the breakdown of the SR approximation for the power spectrum \eqref{eq:PR_from_u} begins already during the SR phase.
    
    \item[T1] The next phase is the \emph{transition from SR to USR}, which we label as T1. This phase is initiated by a feature in the potential that triggers the exit from SR, that is, its onset is signalled by the violation of any of the SR conditions \eqref{eq:SR_cond}. Typically, it is characterized by
    \be
        |\epsilon_3| > 1 \,,
    \ee
    so that at the beginning of this phase, $\epsilon_3=-1$ and $\etah$ starts to increase rapidly. When $\epsh - \etah = 0$, $\epsilon_3$ diverges and changes sign. The phase ends when $\epsilon_3$ decreases to $1$. 
    During this period, $\calH^{-2} z''/z < 0$ has a sharp dip, as can be observed in Fig.~\ref{fig:example_models} in which the T1 phase is indicated by yellow bands. As can be seen in the upper panels of Fig.~\ref{fig:example_models}, the field excursions during this transition tend to be larger than in the following phases. However, as the field is rapidly moving, it's generally much shorter than any other phase.
    
    The variations in model building tend to affect this phase the most. On top of the models considered in Fig.~\ref{fig:example_models}, all of which contain a small local maximum in the potential, this phase can entail rolling through the global minimum~\cite{Yokoyama:1998pt, Saito:2008em, Bugaev:2008bi} or falling down from steep steps in the potential~\cite{Kefala:2020xsx, Inomata:2021uqj, Dalianis:2021iig, Inomata:2021tpx, Cai:2021zsp}. Inflation may end ($\epsh>1$) briefly during this phase (however, it does not in models shown in Fig.~\ref{fig:example_models}).
    
    \item[USR] The \emph{USR period} starts when the field begins to slow down at a plateau or a hillside in the potential (see examples in Fig.~\ref{fig:example_models}). Since $V'\lesssim 0$ and $\epsh \ll 1$, Eq.~\eqref{eq:eom_inf_N} implies that $\partial_N y + 3y \lesssim 0$ and thus $\etah \gtrsim 3$ (see discussion below). Throughout this phase, $\eta_H$ is approximately constant, and $\epsh \propto a^{-2\etah}$ decreases rapidly\footnote{We remark that sometimes the term USR implies a specific growth $\epsh \propto a^{-6}$. Here we do not make this restriction. In particular, the equation $\ddot{\phi} + 3 H \phi \approx 0$ holds only when $\etah = 3$. When $\etah > 3$, the field is not oblivious to the potential and gets slowed down both by Hubble friction and rolling up the potential.}. The USR phase, depicted by a green band in Fig.~\ref{fig:example_models}, is quite brief, typically lasting for a few $e$-folds. The peak of the spectrum consists typically of modes that exited during this phase.
    
    \item[T2] The \emph{dual USR to CR transition}, which we label as T2, is characterized by a sharp decline of $\eta_H$. Like T1, we define it via the condition $|\epsilon_3| > 1$. However, unlike in T1, now also always $\epsh \ll 1$. In Fig.~\ref{fig:example_models}, this phase is denoted by a light green band. One can observe that $\calH^{-2} z''/z$ remains constant throughout this phase---because the background evolution enters the MS equation \eqref{eq:MS} only through this quantity, the evolution of the $u_k$ modes is unaffected by this transition, and no discernible features are generated in the spectrum during this period.
    
    \item[CR] The final \emph{CR period} begins when $|\epsilon_3| < 1$. It is characterized by an almost constant and negative second SR parameter, $\etah<0$. Note that if $|\etah| \ll 1$, this can be considered a second period of SR inflation. Nevertheless, we will use the label CR throughout the article. Note that $\calH^{-2} z''/z$ stays constant throughout the USR, T2 and CR phases, indicating that the CR phase is dual to the USR phase. Since the curvature power spectrum during this phase can be estimated analytically, one can use the duality to infer parts of the spectrum in the two earlier phases. In models considered in the literature (\eg, Model I and II in Fig.~\ref{fig:example_models}), this CR phase often lasts until the end of inflation. However, there can be another phase of SR inflation after the CR phase (\eg, Model III in Fig.~\ref{fig:example_models}). 

\end{description}

The potentials, the power spectra and the evolution of the background quantities $\calH^{-2} z''/z$, $\epsh$, and $\etah$ for three example models are shown in Fig.~\ref{fig:example_models} (for details, see appendix~\ref{app:examples}). We highlight the regions T1, USR, and T2 by yellow, green and light green bands, respectively (this colour convention will be followed throughout the paper).

In all cases, we observe that $\calH^{-2} z''/z$ is constant throughout the USR, T2 and CR phases, implying that these phases are dual~\cite{Wands:1998yp} and relating the spectra produced during the USR and CR periods. To understand this duality better, note that the MS equation depends on the background evolution only via the quantity $z''/z$. Analogously to Eq.~\eqref{eq:SH_modes}, there are also two independent realizations of $z$, corresponding to two independent expansion histories, that will result in mathematically identical MS equations. In an inflating background, one of these will be exponentially increasing, corresponding to the SR or CR phase, while the other is exponentially decreasing---it will describe a USR like phase that is dual to the CR phase. This simple feature lies at the centre of the theory of peaks in the power spectra from single-field inflation. 

The duality dictates that dual phases share the same solutions for $u_k$. The particular solution $u_k$, coming from the initial conditions, stays the same as it evolves through the dual phases. Let us take a closer look at super-horizon evolution during the USR phase. As these modes have exited the horizon, $u_k$ is a superposition of $u^{(1)}_0$ and $u^{(2)}_0$ given Eq.~\eqref{eq:SH_modes} and does not necessarily have to follow $z$ which is temporarily decaying during the USR phase. If the $u_k$ follows $u^{(2)}_0$, then the curvature perturbation $\mc{R}_{c\, k} =u_k/z$ may be enhanced. We will return to this point later (\eg, see Fig.~\eqref{fig:mode_evolution}).

On a more quantitative note, consider the behaviour of the field as it rolls over a small maximum in the potential during the USR+T2+CR phases. To a good approximation, we can assume that $\epsh\ll1$. Expanding the potential around the maximum $\phi = \phi_{\rm m}$, the background evolves as~\eqref{eq:eom_inf_N}
\be \label{eq:phi_max_eom}
    \partial^2_N\phi + 3\partial_N\phi + 3\etav(\phi_{\rm m})\qty(\phi-\phi_{\rm m}) \approx 0 \, ,
\ee
where $\etav(\phi_m)$ is the second potential SR parameter at the maximum. This term is typically of the same order as the first two and cannot be neglected in general. The field thus evolves as 
\be \label{eq:phi_max_solution}
    \phi - \phi_{\rm m} = a^{-3/2} \left( \phi_+ e^{\lambda N} + \phi_- e^{- \lambda N} \right) \, , \qquad
    \lambda \equiv \frac{3}{2}\sqrt{1-\frac{4}{3}\etav(\phi_{\rm m})}\, ,
\ee
where $\phi_{\pm}$ are constant coefficients. Since $\etav(\phi_{\rm m}) \leq 0$ at a maximum, we have $\lambda \geq 3/2$. The `$-$' branch describes the USR phase with $\epsh \propto a^{3 - 2\lambda}$ and the `$+$' branch the CR phase with $\epsh \propto a^{3 + 2\lambda}$. Importantly, ignoring small changes in $H$, we find that
\be \label{eq:zpp_max_solution}
    \calH^{-2} z''/z =  \lambda^2 - 1/4
\ee
is constant as the field rolls over the maximum, independently of $\phi_{\pm}$. This indicates the duality between the corresponding USR and CR phases (and that the T2 phase does not need any special attention). For instance, $\etav(\phi_{\rm m}) \ll 0$ gives the usual SR and the most typical USR behaviour ($\epsh \propto a^{-6}$ or, equivalently, $\ddot{\phi} + 3 H \dot{\phi} \approx 0$), indicating that that the dual of SR is USR with $\epsh \propto a^{-6}$. This behaviour is expected if the PBH-generating feature in the potential is not a maximum, \eg, the USR phase during which the field slows down takes place on a plateau with a negligible slope and curvature.

We stress that all the listed phases are present in single-field models for PBHs considered here, and, strictly speaking, one should not credit a single phase for the peak in the power spectrum. Although the power spectrum reaches its height for modes that exited during the USR phase in quasi-inflection point models shown in Fig.~\ref{fig:example_models}, the evolution of these modes is critically affected by the previous T1 phase, and the spectrum is also tightly linked to the following CR phase via the Wands duality. In models with downward steps~\cite{Kefala:2020xsx, Inomata:2021uqj, Dalianis:2021iig, Inomata:2021tpx, Cai:2021zsp}, in which the enhancement is dominantly due to the "fast" rolling of the inflaton at the step, the downward step is preceded by a USR phase, during which the field slows down by Hubble friction, and finally transitions to an SR phase (or, by our current conventions, to a CR phase) dual to the USR. As an exception, in models with multiple features appearing in rapid succession, the phases listed above may overlap or interact with each other in a non-trivial way~\cite{Tasinato:2020vdk} leading to qualitative differences from the scenarios considered here.

\section{The instantaneous transition approximation}
\label{sec:insta}

Typical models of single-field inflation producing peaks in the power spectrum exhibit a rapid transition from the initial SR phase to the subsequent USR like phase, as is also illustrated in Fig.~\ref{fig:example_models}. The approximation in which this transition is instantaneous can be described by the ansatz
\bea\label{eq:insta}
	\frac{1}{\calH^2}\frac{z''}{z} 
	= -\mathcal{A} \delta(N - N_{c}) +
\left\{
\begin{array}{ll}
	 \lambda_1^2 - 1/4, & \quad N < N_{c} \\
	  \lambda_2^2 - 1/4, & \quad N > N_{c}
\end{array}	\right. \, ,
\eea
where $N_{c}$ is the moment of the transition and $\mathcal{A}$ is a positive constant. The delta peak is added to describe the `dip' in $\calH^{-2} z''/z$ visible in the examples of Fig.~\ref{fig:example_models}. We can shift the $N$ and set $N_{c} = 0$. Informed by Eq.~\eqref{eq:zpp_max_solution}, we included the terms $-1/4$ to simplify later expressions. Imposing continuity, the ansatz \eqref{eq:insta} implies that
\be\label{eq:z_toy}
	z = \left\{
\begin{array}{ll}
	 \zeta_1 e^{(\lambda_1 -\frac{1}{2})(N - N_c)} \, , &  \quad N < N_{c} \\
	 \zeta_2 e^{(\lambda_2 -\frac{1}{2})(N - N_c)} + (\zeta_1-\zeta_2) e^{(-\lambda_2 -\frac{1}{2})(N - N_c)} \, , &  \quad N > N_{c}
\end{array}	\right. \, ,
\ee
with the coefficients $\zeta_i$ related by $\mathcal{A} = \lambda_1 + \left(1 - 2 \zeta_2/\zeta_1\right) \lambda_2$. We assumed that the geometry is almost de Sitter during inflation and took $H$ to be constant. As we will show explicitly below, $\lambda=3/2$ leads to exact de Sitter behaviour with a scale-invariant power spectrum, while $\lambda > 3/2$ ($\lambda < 3/2$) correspond to red-tilted (blue-tilted) power spectra.

The ansatz \eqref{eq:insta} can be intuitively understood via mechanical analogues: The parameter $z$ is initially exponentially growing until it receives a kick with strength $\mathcal{A}$ that slows it down. Strong kicks can invert the velocity and produce a period in which the decaying $z$ mode can dominate, that is, USR. Furthermore, since \eqref{eq:insta} appears in the MS equation~\eqref{eq:MS}, similar reasoning holds for all $u_k$ modes. The main difference is that when $k$ is large enough, the modes will oscillate instead of growing or decaying, weakening the kick's effect on their amplitude.

\begin{figure}[t]
    \centering
    \includegraphics{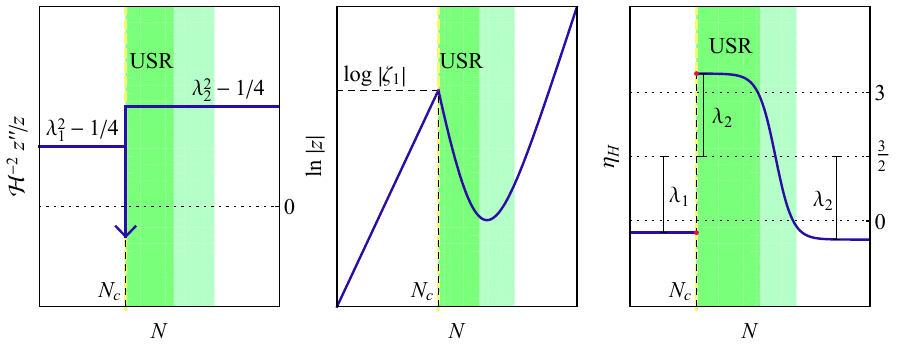}
    \caption{Sketches of $\calH^{-2}z''/z$ from \eqref{eq:insta}, $z$ from \eqref{eq:z_toy}, and $\etah \approx 1-\partial_N\ln|z|$ in the instantaneous T1 transition model. Note that $\calH^{-2}z''/z$ stays constant throughout USR and afterwards, even though $z$ and $\etah$ exhibit non-trivial features. The colour coding of vertical bands is the same as in Fig.~\ref{fig:example_models}.}
    \label{fig:delta_sketches}
\end{figure}

To quantify the inflationary evolution during the different epochs, we compute the second SR parameter, $\etah \approx 1 - \partial_N \ln |z| $, where we neglected $\epsh\approx 0$. To a good approximation, USR transitions to CR when $\etah = 1$ (or equivalently $\partial_N z=0$), thus the duration of the USR phase is
\be \label{eq:N_USR}
	\Delta N_{\rm USR}
	\approx \frac{1}{2 \lambda_2} \ln\qty[\frac{\lambda_2 + 1/2}{\lambda_2-1/2}\qty(\frac{\zeta_1}{\zeta_2}-1)] \, .
\ee
A long USR period with a boosted power spectrum corresponds to $|\zeta_2| \ll |\zeta_1|$ implying that $\mathcal{A}$ in Eq.~\eqref{eq:insta} is tuned to lie slightly below $\lambda_1 + \lambda_2$. Outside the transitions, different terms in \eqref{eq:z_toy} dominate, and we obtain CR inflation with\footnote{
The fact that the Universe passes through different CR phases according to Eq.~\eqref{eq:etaH_regions} can be used to construct an ansatz for $\etah$ for which the curvature power spectrum can be analytically obtained~\cite{Byrnes:2018txb}. An ansatz on $\etah$ instead on $z''/z$, however, will fail to capture the Wands duality between the last two phases and thus it will introduce additional numerical artefacts into the power spectrum. As we show in the following, the ansatz \eqref{eq:insta} of $z''/z$ results from a specific potential and the modulation of the curvature power spectrum can, in our case, be traced back to sharp jumps in the first derivative of this potential.}
\be \label{eq:etaH_regions}
    \etah \approx \left\{
    \begin{array}{lll}
	 \frac{3}{2} - \lambda_1 \, , &  \quad N < N_{c} & \qquad \text{(SR phase)} \\
	 \frac{3}{2} + \lambda_2 \, , & \quad N_{c} < N < N_c + \Delta N_{\rm USR} & \qquad \text{(USR phase)} \\
	 \frac{3}{2} - \lambda_2 \, , & \quad  N \gg N_c + \Delta N_{\rm USR} & \qquad \text{(CR phase)}
    \end{array}	
    \right. \, , 
\ee
as illustrated in Fig.~\ref{fig:delta_sketches}. In the typical scenarios discussed above (see Fig.~\ref{fig:example_models}), we have $\lambda_1 \approx 3/2$, leading to an initial SR phase with $\etah\approx 0$, and $\lambda_2 \gtrsim 3/2$, leading to USR with $\etah \gtrsim 3$ and a final CR phase with $\eta_H \lesssim 0$.

The ansatz \eqref{eq:insta} guarantees the Wands duality between the intermediary USR regime and the final CR. It has the following consequences: First, the respective second SR parameters satisfy
\be
    \eta_{H, \rm USR} + \eta_{H, \text{CR}} = 3 \, ,
\ee
as can be seen from \eqref{eq:etaH_regions}. This follows from the fact that $\calH^{-2} z''/z$ is the same for both phases, as shown in Fig.~\ref{fig:delta_sketches}, linking the two $z$ modes in the $N > N_{c}$ region. Second, since the MS equation~\eqref{eq:MS} depends only on the combination $\calH^{-2} z''/z$, this has important implications for the curvature power spectrum---the spectral properties of curvature fluctuations generated during the USR and the subsequent CR phase are nearly identical. As stated above, the duality is not limited to the simplified model \eqref{eq:z_toy}, but applies to all models in which the peak in the power spectrum is generated when the inflaton rolls over a plateau or a smooth local maximum of its potential, as illustrated in Fig.~\ref{fig:example_models}. The main simplification in the ansatz \eqref{eq:insta} lies in the introduction of the delta function that enforces an instantaneous SR to USR transition.

We now move to solve the MS equation~\eqref{eq:MS} with the ansatz~\eqref{eq:insta}. The general solution is
\be\label{eq:uk}
	u_{k} 
	= 
\left\{\begin{array}{ll}	
	\frac{ \sqrt{\pi}}{\sqrt{\calH} 2\sin(\pi \lambda_1)} \left[ c_{1+} J_{-\lambda_1} \left(k/\calH\right) + c_{1-} J_{\lambda_1} \left(k/\calH\right) \right] \, , &  \quad N < N_{c} \\
	\frac{ \sqrt{\pi}}{\sqrt{\calH} 2\sin(\pi \lambda_2)} \left[ c_{2+} J_{-\lambda_2} \left(k/\calH\right) + c_{2-} J_{\lambda_2} \left(k/\calH \right) \right] \, , &  \quad N > N_{c} \\
\end{array}\right. \, ,
\ee
where $J_\lambda$ is a Bessel function of the first kind. The Bunch-Davies vacuum implies $|c_{1+}| = 1$, $c_{1-} = -e^{-i \pi \lambda_1}c_{1+}$, and the coefficients $c_{2\pm}$ are determined by demanding continuity of $u_k$ at $N_c$ and setting $u_k'$ to jump there as determined by the delta function in \eqref{eq:insta}. The coefficients can be expressed as
\bea\label{c2PlusDelta}
	\frac{c_{2\pm}}{\zeta_2}	
&	=  \left.\mp \frac{\pi \lambda_2}{\zeta_1}  J_{\pm\lambda_2} H_{\lambda_1}
	+ \frac{\pi}{\zeta_2} \frac{k}{2 \calH_c}\left( \pm J_{\pm\lambda_2} H_{\lambda_1-1}  + J_{\pm\lambda_2\pm 1} H_{\lambda_1} \right)\right|_{N = N_c} \, ,
\eea
where we defined $\calH_c \equiv \calH(N_c)$, omitted the argument $k/\calH_c$ of the Bessel and Hankel functions $J_{\lambda}$ and $H_{\lambda}$, and neglected an unimportant phase.

The power spectrum is computed at super-horizon scales where $k \ll \calH$ so the Bessel functions in \eqref{eq:uk} can be approximated with their asymptotic forms. Thus, only the coefficient of the growing mode $c_{2+}$ enters and the power spectrum reads
\bea\label{eq:PR_delta}
	\PR 
& = \frac{k^3}{2\pi^2} |\mc{R}_c|^2	= \frac{k^2 \Gamma(\lambda_2)^2}{4\pi^3} \left| \frac{c_{2+}}{\zeta_2}\right|^2 \left(\frac{k}{2 H e^{N_c}}\right)^{1 - 2 \lambda_2} \\
&	= \frac{k^2}{4\pi} 
	\Bigg| 
	-\frac{\Gamma(1+\lambda_2)}{\zeta_1 } \left(\frac{k}{2 \calH_c}\right)^{-\lambda_2 + 1/2} J_{\lambda_2} H_{\lambda_1}    \\
&   \qquad \quad
    + 	\frac{\Gamma(\lambda_2)}{\zeta_2} \left(\frac{k}{2 \calH_c}\right)^{-\lambda_2+3/2} \left( J_{\lambda_2} H_{\lambda_1-1}  + J_{\lambda_2+1} H_{\lambda_1}\right)  \Bigg|_{N = N_c}^2\,.
\eea
The first bracketed term is rapidly damped as $k^{-\lambda_2-1/2}$ when $k\gg \calH_c$ and thus it contributes mainly to the $k\ll \calH_c$ power spectrum generated during the initial SR phase, indicating that the $\PR$ peak comes from the second term. Its asymptotic behaviour there gives the slope of the power spectrum during SR. Analogously, the asymptotic limits $k \gg \calH_c$ and $k \ll \calH_c$ in the second term give the slopes of the peak away from the top. The dominant contributions to the slopes of the power spectrum can thus be summarized as
\be \label{eq:delta_spectrum_slopes}
    \PR \propto
\left\{\begin{array}{ll}
    k^{3 - 2\lambda_1} \, ,      & \quad k \lesssim \calH_c \\
    k^{5 - 2|\lambda_1-1|} \, ,  & \quad \calH_c \lesssim k \lesssim \calH_c e^{N_{\rm USR}} \\
    k^{3 - 2\lambda_2} \, ,      & \quad \calH_c e^{N_{\rm USR}} \lesssim k
\end{array}\right. \, .
\ee
\begin{figure}[t]
    \centering
    \includegraphics{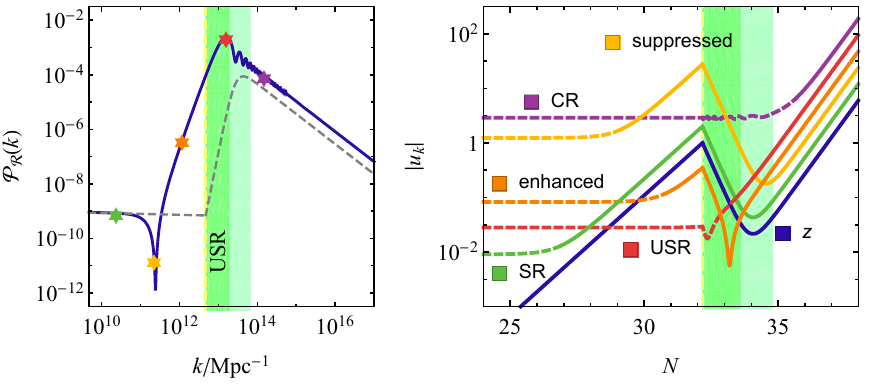}
    \caption{\emph{Left panel:} The curvature power spectrum \eqref{eq:PR_delta} in the instantaneous transition approximation for the benchmark model \#1 from table~\ref{tab:delta_models} (blue) and, for comparison, the SR approximation $\PR = H^2/(8\pi^2 \mpl^2\epsh)$ with $\epsh=z^2/(2\mpl^2a^2)$ (dashed grey), with a number of modes from different regions highlighted. \emph{Right panel:} Evolution of the Sasaki-Mukhanov variable $u_k$ for the corresponding wave numbers from \eqref{eq:uk}. The dashed and solid sections correspond to the sub- and super-horizon regions, $\calH < k$ and $\calH > k$, respectively. The modes are normalized so that their large $N$ asymptotic is the same order as $z$, which in turn is normalized as $z(N_c)=1$. The colour coding of vertical bands is the same as in Fig.~\ref{fig:example_models}.}
    \label{fig:mode_evolution}
\end{figure}
For small and large $k$, \eqref{eq:delta_spectrum_slopes} gives a scale invariant spectrum if $\lambda_{1,2}=3/2$, corresponding to a pure de Sitter case, while $\lambda_{1,2} > 3/2$ ($\lambda_{1,2} < 3/2$) will produce a red-tilted (blue-tilted) power spectrum. The peak's rising slope is sensitive to the behaviour at $N < N_c$ through $\lambda_1$. As long as $\eta_{H}<0$ during the SR phase, that is, $\lambda_1 > 3/2$ from \eqref{eq:etaH_regions}, the growth exponent has a maximal value of $4$, consistent with previous studies~\cite{Byrnes:2018txb,Carrilho:2019oqg}. In case the spectrum is blue-tilted during the initial phase, \ie, $1 < \lambda_1 < 3/2$, the power spectrum can grow as fast as $\PR \propto k^{5}$, which is similar to the maximal growth $\PR \propto k^{5}\log^2k$ found in Ref.~\cite{Carrilho:2019oqg}.

The position and height of the maximum of the power spectrum \eqref{eq:PR_delta} are given by\footnote{A decent numerical fit is given by $k_{\rm max}/\calH_c = 2.1 + 0.74 \lambda_2-0.13 \lambda_2 \ln \lambda_2$. It is accurate within 1\% in the range $3/2<\lambda_1 < 2$, $3/2<\lambda_2 < 30$.}
\be
    k_{\rm max} = \mathcal{O}(3) \calH_c \, , \qquad
    \PR(k_{\rm max}) = \mathcal{O}(0.05)/\zeta_2^2\, ,
\ee 
respectively. This estimate has only a mild dependence on $\lambda_2$, and is almost independent of $\lambda_1$. Importantly, note that the position of the maximum depends only on the comoving horizon scale $\calH_c$ at the T1 phase, while the height of the peak depends on $\zeta_2$ and thus, by Eq.~\eqref{eq:N_USR}, on the duration of USR. This is expected, because the features of the peak in the spectrum \eqref{eq:PR_delta} arise from the coefficient $c_+$ determined by matching of the modes at $N_c$, that is, the shape of the peak is determined by mode evolution during T1 phase.

An example of the power spectrum \eqref{eq:PR_delta} is displayed in Fig.~\ref{fig:mode_evolution}. The right panel shows the evolution of the $u_k$ modes exiting the horizon at different times. Modes that exit much before or after the USR freeze almost immediately after horizon exit. That is, they follow $z$ so that $\mathcal{R} = u_k/z = {\rm const.}$ (see the discussion in sections \ref{sec:perturbation_evolution} and \ref{sec:phases}). Modes that exit during or close to USR deviate from $z$ for some time after horizon crossing, briefly following the second solution $u^{(2)}_0$ in \eqref{eq:SH_modes}, which dominates during USR. At $N_c$, we can see the kicks due to the delta function in~\eqref{eq:insta}. Both $u_k$ and $z$ have a tendency to grow, and the growth of the curvature perturbation $|\mc{R}_{c\, k}| =\left\lvert u_k/z\right\rvert$ during USR is due to the decrease in $z$ rather than the growth of $u_k$. Indeed, all the modes depicted in Fig.~\ref{fig:mode_evolution} (except the CR mode that exits well after USR) decrease at least momentarily during USR.

Note the super-horizon enhancement of modes that exit a few $e$-folds before the USR phase, at a time when the SR conditions still hold. The power spectrum in the SR approximation \eqref{eq:PR_from_u} is shown in the left panel of Fig.~\ref{fig:mode_evolution} for comparison---while the SR approximation would predict a flat spectrum before the USR phase, the MS equation shows rapid growth. The power spectrum is also mildly enhanced during the final CR phase when compared to the naive SR expectation.

\subsection{Potentials producing instantaneous SR to USR transitions} 
\label{sec:delta_potential}

\begin{figure}[t]
    \centering
    \includegraphics[width=0.65\textwidth]{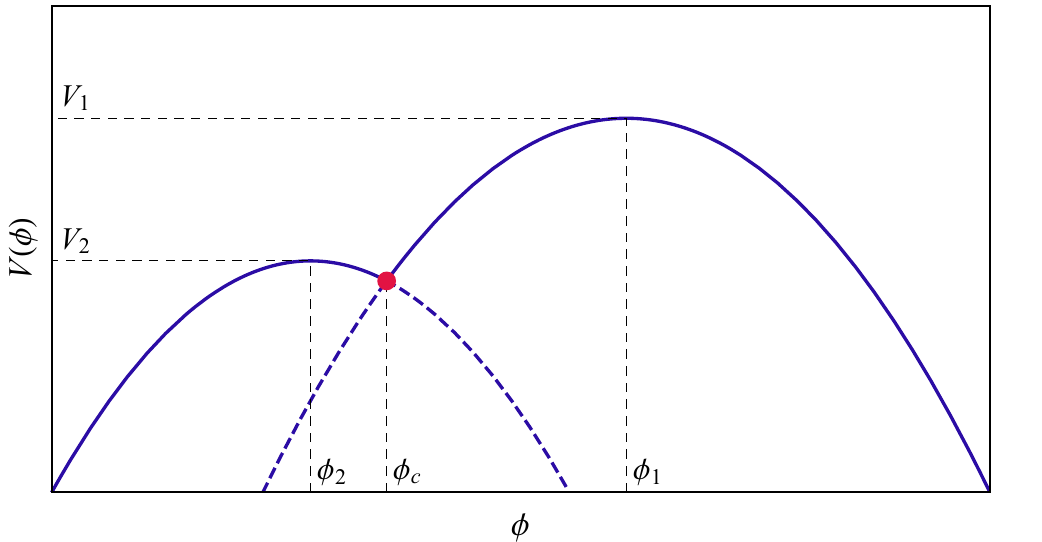}
    \caption{A sketch of the instantaneous transition model potential \eqref{eq:V_humps}.}
    \label{fig:delta_V_sketch}
\end{figure}

As we will show in this section, the behaviour \eqref{eq:insta} of $\calH^{-2}z''/z$ can be realized by a scalar field rolling down a potential consisting of two concave parabolas, sketched in Fig.~\ref{fig:delta_V_sketch},
\be 
\label{eq:V_humps}
    V = \left\{
    \begin{array}{ll}
     V_2\qty(1+\frac{1}{2\mpl^2}\eta_{V2}\qty(\phi-\phi_2)^2) \, , & \quad \phi < \phi_c \phantom{\bigg|}
	 \\
	 V_1\qty(1+\frac{1}{2\mpl^2}\eta_{V1}\qty(\phi-\phi_1)^2) \, , &  \quad \phi > \phi_c \phantom{\bigg|}
    \end{array}	
    \right. \, .
\ee
Here $\eta_{V1},\eta_{V2} < 0$ are the potential SR parameters $\eta_V$ at the two hilltops at $\phi_1$ and $\phi_2$. We set $\phi_2 < \phi_c < \phi_1$ and choose $V_1 > V_2$. The transition point $\phi_c$ is at the intersection of the two parabolas so that the potential is continuous. We further choose the parameters so that $V(0)=0$.

During its evolution, the field starts close to the higher hilltop, at $\phi \lesssim \phi_1$, rolls down to the sharp feature at $\phi_c$, and continues rolling along the other parabola, crossing the lower hilltop at $\phi_2$ and continuing towards $\phi=0$. We require $\epsilon_V \ll 1$ during this evolution; this is possible if both of the hilltops are sufficiently close to $\phi_c$. In this limit, $\epsh \ll 1$ for the full duration, and $H$ is approximately a constant, say, $H \approx\sqrt{V_2/3}/\mpl$, as is $\eta_V$ separately on the two parabolas. We will use these approximations repeatedly throughout this section. We also take $|\eta_{V1}| < 1$, so that the field starts in SR. The SR approximation is only broken at $\phi_c$, where $\eta_V$ has a negative delta spike. Comparing this to \eqref{eq:zpp} and using all the above-mentioned approximations, including $\epsh \ll 1$, we see that $z''/z$ matches the approximation \eqref{eq:insta} with\footnote{Note that in SR, $\eta_V \approx \eta_H \ll 1$, this coincides with the $\lambda_i \approx 3/2-\eta_H$ from \eqref{eq:etaH_regions}.}
\be 
\label{eq:lambda_in_etaV}
    \lambda_i \approx \frac{3}{2}\sqrt{1-\frac{4}{3}\eta_{Vi}} \, , \qquad i=1,2 \, ,
\ee
consistent with Eq.~\eqref{eq:phi_max_solution}.

Let us next study the field's evolution more closely. In the process, we will make contact with the rest of the formalism presented in section \ref{sec:insta}.

Near the first hilltop, the field first traverses from an initial value $\phi_*$ to the transition point $\phi_c$ in SR. The duration of this phase in $e$-folds is approximately
\be \label{eq:N_SR_initial}
    N \approx \int_{\phi_c}^{\phi_*} \frac{\dd \phi}{\mpl\sqrt{2\eps_V}} \approx \frac{1}{|\eta_{V1}|} \ln \frac{\Delta \phi_1}{\phi_1 - \phi_*} \, , 
\ee
where we defined $\Delta \phi_i \equiv |\phi_i - \phi_c|$. After SR ends, the field continues to evolve in the general way outlined in section \ref{sec:phases}. Analogously to the Eq.~\eqref{eq:phi_max_eom}, the field obeys
\be \label{eq:phi_humps_eom}
    \partial^2_N\phi + 3\partial_N\phi + 3\eta_{V2}\qty(\phi-\phi_2) = 0 \, ,
\ee
and thus
\be 
\label{eq:phi_humps_solution}
    \phi - \phi_2 = c_+ e^{(-\frac{3}{2} + \lambda_2)\qty(N-N_c)} + c_- e^{(-\frac{3}{2} - \lambda_2)\qty(N-N_c)} \, ,
\ee
where we used \eqref{eq:lambda_in_etaV}. Note that $z=a\partial_N\phi$, and \eqref{eq:phi_humps_solution} is indeed consistent with the $z$ approximation \eqref{eq:z_toy}. The coefficients $c_\pm$ are fixed by demanding that $\phi=\phi_c$ and $N=N_c$ and that the field velocity is continuous. Using the approximation $\partial_N\phi_c \approx -\mpl\sqrt{2\epsilon_V} \approx \eta_{V1}\Delta\phi_1$ on the SR side\footnote{If one wants to fine-tune the evolution and thus the height of the power spectrum peak to a great degree, it may be necessary to include higher-order SR corrections to $\partial_N\phi_c$.}, we find
\be \label{eq:c_plus_minus}
    c_\pm = \frac{1}{2\lambda_2} \left(\mp|\partial_N\phi_c| + \Delta\phi_2 (\pm 3/2 + \lambda_2)\right)\, .
\ee
Comparing \eqref{eq:phi_humps_solution} to the $z$ approximation \eqref{eq:z_toy} then gives
\be \label{eq:match_c_expressions}
    \frac{\zeta_2}{\zeta_1} 
    = \frac{\lambda_2-3/2}{2\lambda_2}\qty(1-(3/2+\lambda_2)\frac{\Delta \phi_2}{|\partial_N\phi_c|}) \, .
\ee
Eqs.~\eqref{eq:lambda_in_etaV} and \eqref{eq:match_c_expressions} relate the instantaneous transition ansatz \eqref{eq:z_toy} for $z$ to the piecewise smooth potential \eqref{eq:V_humps}. This correspondence holds up to shifts in $\phi$. The height of the resulting power spectrum $\PR$ can be changed by changing the height of the potential and thus $H$. This corresponds to rescaling $\zeta_1$ in \eqref{eq:PR_delta}.

\begin{figure}[t]
    \centering
    \includegraphics[width=0.98\textwidth]{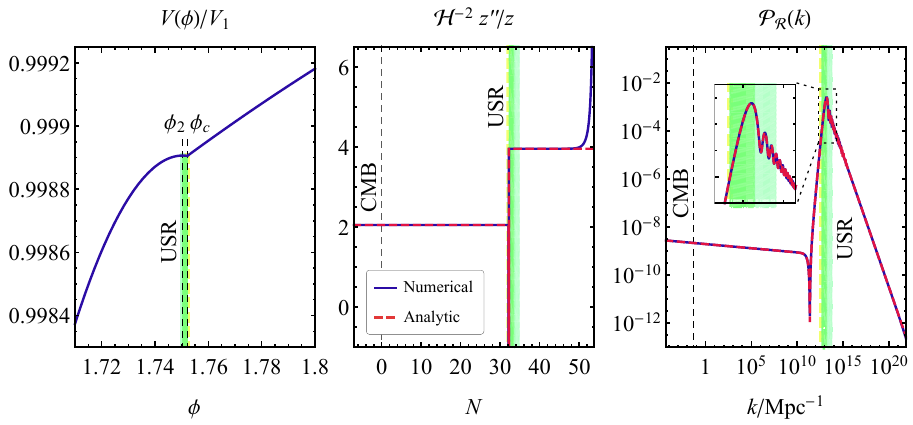}
    \caption{The potential {\it (left panel)}, $\calH^{-2}z''/z$ from \eqref{eq:zpp} {\it (middle panel)}, and the power spectrum {\it (right panel)} for the potential~\eqref{eq:V_humps} with the parameter values from table~\ref{tab:delta_models} (model \#1). The blue curves correspond to the full numerical solution with the potential~\eqref{eq:V_humps}, whereas the red, dashed curves correspond to the analytical approximations~\eqref{eq:insta}, \eqref{eq:PR_delta} with parameters determined from~\eqref{eq:lambda_in_etaV}, \eqref{eq:match_c_expressions}. The colour coding of vertical bands is the same as in Fig.~\ref{fig:example_models}, and the CMB scales have been marked where appropriate. Notice that the potential is shown only close to the feature at $\phi_c$. In the middle panel, the $x$-axis marks the number of $e$-folds from the CMB. The inset highlights that the potential~\eqref{eq:V_humps} perfectly reproduces the oscillations at the $\PR$ peak found in the approximation~\eqref{eq:PR_delta}.}
    \label{fig:delta_row}
\end{figure}

In the limit of small $\eta_{V2}$, the length of the USR period \eqref{eq:N_USR} can be written to leading order as
\be 
\label{eq:N_USR_2}
    \Delta N_{\rm USR} 
    \approx \frac{1}{3}\ln \frac{6}{-\eta_{V2}\qty(1-\frac{3\Delta\phi_2}{|\partial_N\phi_c|})} \, .
\ee
Substituting this to \eqref{eq:phi_humps_solution}, the $\phi$-value at the end of USR is, to leading order,
\be 
\label{eq:phi_int_approx}
    \phi_{\rm CR} \approx \phi_2 + c_+ \approx \phi_c - |\partial_N \phi_c|/3 \, .
\ee
Since $c_+ < 0$, this is located at a field value somewhat smaller than the lower hilltop $\phi_2$. At this level of approximation, the result agrees with pure USR, where the potential term in \eqref{eq:phi_humps_eom} is neglected and the field is left to roll on a plateau until it stops after traversing a field range of $|\partial_N\phi_c|/3$. If $\eta_{V2}$ is not small, $\phi_{\rm CR}$ can still be solved similarly, but the expressions become lengthy.

The USR phase transitions to CR, where it stays until inflation ends close to $\phi=0$. The remaining number of $e$-folds is approximately~\cite{Martin:2012pe}
\bea
\label{eq:N_SR_final}
    \Delta N_{\rm CR} 
    &\approx (1-\etah/3)\int_0^{\phi_{\rm CR}}\frac{\dd \phi}{\mpl\sqrt{2\epsilon_V}} \\
    &= \qty(\frac{1}{2}+\frac{1}{3}\lambda_2)\qty[ - \frac{1}{\eta_{V2}}\ln \frac{\phi_2}{\phi_2 - \phi_{\rm CR}} - \frac{\phi_2^2 - (\phi_2 - \phi_{\rm CR})^2}{4 \mpl^2}] \, ,
\eea
where we dropped the constant $H$ approximation, which breaks down after the field leaves the lower hilltop, but retained the CR assumption of a constant $\etah$. This concludes our description of the inflaton's time evolution.

The potential \eqref{eq:V_humps} can be considered a generalization of the Starobinsky model~\cite{Starobinsky:1992ts}, built out of two linear sections with different slopes. Both models exhibit a similar discontinuity in the first derivative of the potential. Indeed, if we set $\phi_2 > \phi_c$, so that the potential grows monotonously all the way to $\phi_1$, the resulting dynamics would be similar to the Starobinsky case near $\phi_c$. However, Starobinsky's scenario cannot support a prolonged period of USR. For that, we need $\phi_2 < \phi_c$ so that the potential has a secondary maximum, slowing the field down as it rolls over the hilltop. We remark that analogous simplified models involving piecewise quadratic potentials have been used to construct analytic descriptions of preheating in plateau inflation~\cite{Koivunen:2022mem}. 

\subsection{Constructing potentials for PBHs of any mass}
\label{sec:delta_example}

Using the results of the previous sections, we can now tailor a potential to produce PBHs with a desired mass and abundance, simultaneously with desired CMB observables. The full SR parameters of the potential \eqref{eq:V_humps} read
\be 
\label{eq:hump_SR_parameters}
    \epsilon_V = \frac{\frac{1}{2\mpl^2}\qty(\phi-\phi_i)^2\eta_{Vi}^2}{\qty(\frac{1}{2\mpl^2}\qty(\phi-\phi_i)^2\eta_{Vi} + 1)^2}\, , \qquad
    \eta_V = \frac{\eta_{Vi}}{\frac{1}{2\mpl^2}\qty(\phi-\phi_i)^2\eta_{Vi} + 1} \, ,
\ee
where $i=1$ for $\phi>\phi_c$ and $i=2$ for $\phi<\phi_c$. Starting from the desired values for $n_s$ and $r$, presented in terms of the SR variables in \eqref{eq:SR_ns_and_r}, we can solve $\eta_{V1}$ and $\phi_{\rm CMB} - \phi_1$ from \eqref{eq:hump_SR_parameters}, where $\phi_{\rm CMB}$ is the CMB field value. Fixing $A_s$ then sets $V_1$ through \eqref{eq:PR_from_u} using $H^2 = V/(3\mpl^2)$ and $\epsh\approx\epsv$. The desired PBH mass gives the position of the power spectrum peak from \eqref{eq:M_k} below with $\Delta N \approx \Delta\ln k$ as $e$-folds from the CMB. The peak will be located roughly at the transition scale $N=N_c$, so Eq.~\eqref{eq:N_SR_initial} with $\phi_*=\phi_{\rm CMB}$ sets $\phi_c$. The parameters $\eta_{V2}$ and $\Delta\phi_2$ can then be used to adjust the height of the power spectrum peak and the remaining number of $e$-folds of inflation, setting the PBH abundance and the total length of inflation. The length of USR \eqref{eq:N_USR_2} can be a useful proxy for the abundance, and is mainly sensitive to $\Delta\phi_2$. In turn, $\eta_{V2}$ controls the remaining number of $e$-folds, approximated by \eqref{eq:N_SR_final}. The remaining parameters in \eqref{eq:V_humps}, that is, $V_2$ and $\phi_2$, are fixed by the requirements of continuity at $\phi_c$ and the requirement $V(0)=0$.

\begin{table}
\begin{center}
\begin{tabular}{c@{\hskip 20pt}cccc@{\hskip 20pt}cc}
\toprule
Model & i & $V_i / M_P^4$ & $\eta_{Vi}$ & $\phi_i$ & $\phi_c$ & $\phi_\mathrm{CMB}$ \\
\midrule
\#1 & 1 & $3.1100369\times10^{-12} $ & $-0.017475$ & $2.10607$ & $1.75215$ & $1.9038$ \\
& 2 & $3.1066362\times10^{-12}$ & $-0.652743$ & $1.75043$ & & \\
\#2 & 1 & $3.1100369\times10^{-12} $ & $-0.017475$ & $2.62987$ & $2.35685$ & $2.4276$ \\
& 2 & $3.1080125\times10^{-12}$ & $-0.360486$ & $2.35543$ & & \\
\bottomrule
\end{tabular}

\vspace{1cm}

\begin{tabular}{cccccccc}
\toprule
Model & $n_s$ & $r$ & $N_\text{CMB}$ & $k_\text{peak}$ (Mpc$^{-1}$) & $\PR{}_\text{peak}$ & $M_\text{PBH}$ (g) \\
\midrule
\#1 & $0.965$ & $0.99\times10^{-4}$ & $53.7$ & $1.64\times 10^{13}$ & $0.0025$ & $1.04\times10^{20}$ \\
\#2 & $0.965$ & $0.99\times10^{-4}$ & $55.6$ & $5.35\times 10^{6}$ & $0.014$ & $9.75\times10^{32}$ \\
\bottomrule
\end{tabular}
\end{center}

\caption{Input parameters and the resulting cosmological observables for the potential~\eqref{eq:V_humps} in two example cases, corresponding to PBHs of approximately $10^{20}$g (model \#1) and a solar mass (model \#2). For the input parameters, no digits have been omitted: the observables were computed with these exact values. Tuning $\PR{}_\text{peak}$ more finely requires more accuracy in the inputs. The CMB parameters fit the observational constraints \cite{Planck:2018jri}.}
\label{tab:delta_models}
\end{table}

We carried out this exercise and produced two example models, compatible with the CMB observations \eqref{eq:CMB_observables} and tuned to produce PBHs with masses $10^{20}$ and $10^{33}$g. We used the analytical expressions in section \ref{sec:delta_potential} to fix the CMB and as a starting point for tuning the power spectrum peak, but solved the background evolution numerically starting from the potential \eqref{eq:V_humps}, and numerically optimized the parameters $\Delta\phi_1$, $\Delta\phi_2$, and $\eta_{V2}$.  The observables are listed in table~\ref{tab:delta_models}, given the analytical power spectrum \eqref{eq:PR_delta}. The input parameters for the spectrum were computed from the potential using Eqs.~\eqref{eq:lambda_in_etaV} and \eqref{eq:match_c_expressions}. The resulting parameter values for the potential are listed in table~\ref{tab:delta_models}. We then solved the MS equation~\eqref{eq:MS} numerically for approximately $2000$ modes in model \#1 and compared to the analytical approximation. The results are depicted in Fig.~\ref{fig:delta_row} and show an excellent match between the analytics and the numerics.\footnote{To avoid numerical issues related to the discontinuity in the first derivative of the potential, we used the transformation in Eq.~(3.23) of Ref.~\cite{Rasanen:2018fom}.}

The power spectrum shown in Fig.~\ref{fig:delta_row} is typical for a PBH model and, due to its easy-to-manipulate analytical form and a clear connection to a potential, can be used as a proxy for studying observational signals of such models. The only potentially unrealistic feature of the spectrum are the oscillations near the peak, originating from the Bessel functions in~\eqref{eq:PR_delta}, and induced by the discontinuity at $N_c$. In the next section, we modify the model to reduce the amplitude of these oscillations.

\section{Approximating the spectra of smooth potentials}
\label{sec:spectrum_approximation}

\begin{figure}[t]
    \centering
    \includegraphics{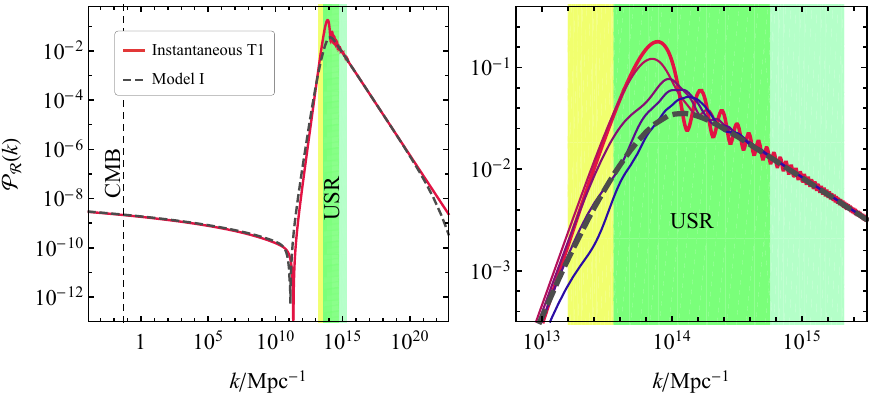}
    \caption{Model I from figure \ref{fig:example_models} compared to the approximation in Eq.~\eqref{eq:PR_delta_approx}. On the right hand side, fits using the non-instantaneous transition model are also plotted with varying durations of the T1 phase, $\Delta N_{\rm T} = \{0.5, 1, 1.5, 2\}$ $e$-folds from top to bottom. The remaining parameters were chosen to match the slopes of the peak. The colour coding of vertical bands indicating different inflationary phases is the same as in Fig.~\ref{fig:example_models}.}
    \label{fig:model_1_fits}
\end{figure}

The instantaneous transition model of the previous section has two shortcomings: first, it produces a spectrum with a constant spectral index in the SR region, which tends to not capture well the spectrum all the way from the CMB to the PBH scale in smooth models such as those depicted in Fig.~\ref{fig:example_models}. Second, it exhibits sharp oscillations in the power spectrum peak, absent in the smooth models. In this section, we attempt to overcome these flaws and approximate the spectra of such smooth models analytically. Fig.~\ref{fig:model_1_fits} shows the resulting example fits.

\subsection{Fitting the instantaneous transition approximation} 
\label{sec:delta_fit}

To apply the instantaneous transition approximation to realistic models of inflation, we will express the spectrum~\eqref{eq:PR_delta} in terms of SR parameters. In the $k\ll \calH_c$ and $k\gg \calH_c e^{N_{\rm USR}}$ asymptotics of Eq.~\eqref{eq:PR_delta} we find that the power spectrum can be approximated by
\be\label{eq:PR_CR}
    \PR^{(\rm CR)}(k; \lambda_i)
    \equiv 
    \left.\frac{\Gamma(\lambda_i)^2}{2^{4-2\lambda_i}\pi^3} \frac{H^2}{\mpl^2 \epsh} \right|_{k=\calH},
\ee

\noindent that is, by the usual CR power spectrum~\cite{Kinney:2005vj, Motohashi:2014ppa, Motohashi:2019rhu}.\footnote{The $k\ll \calH_c$ asymptotics of Bessel functions give
\bea\label{eq:approx_I}
    -\frac{\Gamma(1+\lambda_2)}{\zeta_1 } J_{\lambda_2} H_{\lambda_1} \left(\frac{k}{2 \calH_c}\right)^{-\lambda_2 + 1/2}
&   \approx 
    i \frac{\Gamma(\lambda_1)}{\pi z} \left(\frac{k}{2 \calH}\right)^{-\lambda_1+1/2}\, ,
\eea
where we used \eqref{eq:z_toy} to eliminate $\zeta_1$. In the last expression, $z$ and $\calH$ can be evaluated at an arbitrary time $N < N_c$. We replace $\zeta_{1}$ and $\zeta_{2}$ in Eq.~\eqref{eq:PR_delta} with $\epsh = z^2/(a \mpl)^2/2$ (and $z$ given by Eq.~\eqref{eq:z_toy}) during the SR or CR phases, respectively.} The SR approximation is recovered by $\PR^{(\rm SR)}(k) \equiv \PR^{(\rm CR)}(k; 3/2)$. In this region, the modes stop evolving after horizon exit, so the usual matching $k=\calH$ can be performed. In this way, it is possible to account for an evolving $H$ during the initial SR phase, going beyond the constant $H$ approximation of the previous section. If the $\PR$ peak is due to a small local maximum in the potential, the $\lambda_2$ can be read off from the potential using Eq.~\eqref{eq:lambda_in_etaV}.\footnote{Having the numerical solution at hand, one may also use $\lambda_2 \approx 3/2 - \eta_H$ from Eq.~\eqref{eq:delta_spectrum_slopes} instead of $\etav$ with Eq.~\eqref{eq:lambda_in_etaV}. In the example in Fig.~\ref{fig:model_1_fits}, the difference in the $\lambda_2$ estimate is only 0.1\%.} During the final CR phase, we can fix the normalization of the second term in~\eqref{eq:PR_delta} at some fixed scale $k_{\rm CR}$ from the analytic CR power spectrum~\eqref{eq:PR_CR}. Assuming that the SR approximation is valid during the initial phase, we can choose $\lambda_1 = 3/2$. We can then write down an improved version of \eqref{eq:PR_delta} as
\bea\label{eq:PR_delta_approx}
	\PR 
	\approx  
    \Bigg| \theta(\calH_c - k) \sqrt{\PR^{(\rm SR)}}|_{\calH=k} 
   + \frac{k \, p(k/\calH_c,\lambda_2)}{k_{\rm CR} \, |p(k_{\rm CR}/\calH_c,\lambda_2)|} \sqrt{\PR^{(\rm CR)}(k_{\rm CR};\lambda_2}) \Bigg|^2\, ,
\eea
where
\be
    p(\kappa;\lambda_2)
    \equiv e^{i \kappa}\left(\frac{\kappa}{2}\right)^{1-\lambda_2} 
    \left( 
    -   J_{\lambda_2}\left(\kappa\right)   
    +   \left(i-\frac{1}{\kappa}\right) J_{\lambda_2+1}\left(\kappa\right) \right)
\ee
describes the peak generated in the instantaneous SR to USR transitions (we expanded $H_{\lambda_1}$ and $H_{\lambda_1-1}$ from \eqref{c2PlusDelta} with $\lambda_1=3/2$). The step function $\theta$ in the first term models the $k \gg \calH_c$ damping of the first term in Eq.~\eqref{eq:PR_delta}. As a fit to a realistic model, \eqref{eq:PR_delta_approx} is fully determined by the evolution of background quantities solved from \eqref{eq:eom_inf_N}, in particular $H$ and the SR parameters $\epsh$ and $\etah$, together with the transition point given by $\calH_c$. We remark that Eq.~\eqref{eq:PR_delta_approx} may be improved further by modifying the shape of $p(\kappa;\lambda_2)$.

In Fig.~\ref{fig:model_1_fits}, the approximation \eqref{eq:PR_delta_approx} is compared to the spectrum from a smooth theoretically well-motivated potential---Model I of Fig.~\ref{fig:example_models}, non-minimal quartic inflation (see appendix~\ref{app:examples} for details). For this potential, $\lambda_2 = 1.916$. Apart from the oscillations at the peak, the instantaneous transition approximation \eqref{eq:PR_delta_approx} provides a good model of peaked spectra, fitting both the SR and CR regions well.

In single-field inflation, a degree of tuning is typically required to obtain the desired height of the power spectrum peak. Eq.~\eqref{eq:PR_delta_approx} can help with this by providing an approximate analytic representation of the power spectra. In this case, the tuning of the potential parameters to several significant digits can be circumvented as the duration of $\Delta N_{\rm USR}$, or equivalently, the height of the $\PR$ peak can be set by hand. The rest of the spectrum in \eqref{eq:PR_delta_approx} can be estimated using the SR and CR approximations to obtain $\epsh$. This method is relatively fast and can be useful, especially in inflationary model building or GW phenomenology when the relevant observables (\eg, $r$ and $n_s$) are relatively insensitive to small changes in the height of the $\PR$ peak. On the other hand, the approximation \eqref{eq:PR_delta_approx} does not capture the shape of the peak perfectly and is not suitable for precision estimates of the PBH abundance, which can be sensitive to minor adjustments in the peak. We will return to this issue in section~\ref{sec:PBHs}.

\subsection{Non-instantaneous SR to USR transitions}
\label{sec:non-insta}

\begin{figure}[t]
    \centering
    \includegraphics{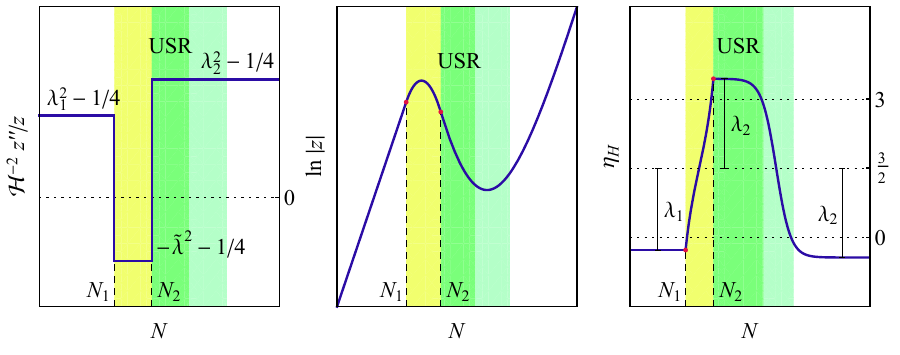}
    \caption{Sketches of $\mathcal{H}^{-2}z''/z$ from \eqref{eq:box}, $z$ from \eqref{eq:z_box}, and $\etah \approx 1-\partial_N\ln|z|$ in the model with non-instantaneous transitions. The colour coding of vertical bands matches Fig.~\ref{fig:example_models}}
    \label{fig:box_sketches}
\end{figure}

Although the instantaneous transition approximation captures many of the features of the peak in $\PR$, it produces oscillations around the peak that are not observed with non-singular inflaton potentials. In the following, we will show that this modulation is an artefact of the infinitely sharp transition and can be softened. Similar oscillations in $\PR$ have been observed when considering inflationary cosmologies with sharp jumps in $\etah$ and, in a recent work, they have been shown to disappear once the jumps are smoothed~\cite{Cole:2022xqc}.

Instead of an instantaneous kick as in Eq.~\eqref{eq:insta}, let us then consider an intermediate T1 phase lasting $\Delta N_{\rm T} \equiv N_2-N_1$ $e$-folds. This scenario can be modelled by the ansatz 
\bea\label{eq:box}
	\calH^{-2}\frac{z''}{z} 
	= 
\left\{
\begin{array}{ll}
	  \lambda_1^2 - 1/4, & \quad N < N_1 \\
	  -\tilde\lambda^2 - 1/4, & \quad N_1 < N < N_2 \\
	  \lambda_2^2 - 1/4, & \quad N > N_2
\end{array}	\right. \, ,
\eea
where the tilde marks quantities during the T1 phase and $\lambda_{1,2}$ have the same interpretation as in the instantaneous case. The background evolution must now be divided into three epochs,
\be\label{eq:z_box}
	z = \left\{
\begin{array}{ll}
    \zeta_1 e^{(\lambda_1 -\frac{1}{2})(N - N_1)}, 
    &  \quad N < N_1,
    \\
	\tilde\zeta_{+} e^{(i\tilde\lambda -\frac{1}{2})(N - N_1)} + \tilde\zeta_{-} e^{(-i\tilde\lambda -\frac{1}{2})(N -N_1)}, 
    &  \quad N_1 < N < N_2,
    \\
	\zeta_{2+} e^{(\lambda_2-\frac{1}{2})(N-N_2)} + \zeta_{2-} e^{(-\lambda_2-\frac{1}{2})(N-N_2)}
    & \quad N_2 < N,
\end{array}	\right.
\ee
where the coefficients $\zeta_i$ are given in appendix~\ref{app:box}. An example of $z$ evolution \eqref{eq:z_box} is shown in Fig.~\ref{fig:box_sketches}. This should be compared to Fig.~\ref{fig:delta_sketches}: the sharp features in the instantaneous transition have been eliminated. Setting the end of USR to the point where $\etah=1$, the duration of the USR phase is
\be\label{eq:N_USR_box}
	\Delta N_{\rm USR}
	\approx \frac{1}{2 \lambda_2} \ln\qty[\frac{\lambda_2 + 1/2}{\lambda_2-1/2} \frac{\zeta_{2-}}{\zeta_{2+}} ] \, .
\ee
Noting that in the instantaneous case $\zeta_{2-} = \zeta_{1} - \zeta_{2+}$, this expression is identical to Eq.~\eqref{eq:N_USR}. An extended USR period is realized when the growing mode is strongly suppressed at the beginning of the USR phase, \ie, $|\zeta_{2+}| \ll |\zeta_{2-}|$.

Analogously, the general solution to the MS equation now reads
\be\label{eq:MSbox}
	u_{k} 
	= 
\left\{\begin{array}{ll}	
	\frac{ \sqrt{\pi}}{\sqrt{\calH} 2\sin(\pi \lambda_1)} \left[ 
	c_{1+} J_{-\lambda_1} \left(\frac{k}{\calH}\right) + 
	c_{1-} J_{\lambda_1} \left(\frac{k}{\calH}\right) \right] \, , 
	&  \quad N < N_1 \\
	\frac{ \sqrt{\pi}}{\sqrt{\calH} 2\sin(\pi i\tilde\lambda)} \left[ 
	\tilde c_{+} J_{-i\tilde\lambda} \left(\frac{k}{\calH}\right) + 
	\tilde c_{-} J_{i\tilde\lambda} \left(\frac{k}{\calH}\right) \right] \, , 
	&  \quad N_1 < N < N_2 \\
	\frac{ \sqrt{\pi}}{\sqrt{\calH} 2\sin(\pi \lambda_2)} \left[ 
	c_{2+} J_{-\lambda_2} \left(\frac{k}{\calH}\right) + 
	c_{2-} J_{\lambda_2} \left(\frac{k}{\calH}\right) \right] \, ,
	&  \quad N_2 < N
\end{array}\right. \, ,
\ee
with the expressions for the coefficients $c_i$ listed in appendix~\ref{app:box}. The power spectrum is again determined from the coefficient $c_{2+}$ of the growing mode,
\be \label{eq:PR_box}
    \PR 
    = \frac{k^3}{2\pi^2} |\mc{R}_c|^2
    = \frac{k^2 \Gamma(\lambda_2)^2}{4\pi^3} \left|\frac{c_{2+}}{\zeta_{2+}}\right|^2 \left(\frac{k}{2 \calH_2} \right)^{1 - 2 \lambda_2} \, ,
\ee
where $\calH_2 \equiv \calH(N_2)$. Due to the complicated form of the expressions, we do not study the details of \eqref{eq:PR_box} further, but we note that the asymptotic behaviour in the SR and CR regions coincides with the instantaneous transition approximation.

\begin{figure}[t]
    \centering
    \includegraphics[width=0.65\textwidth]{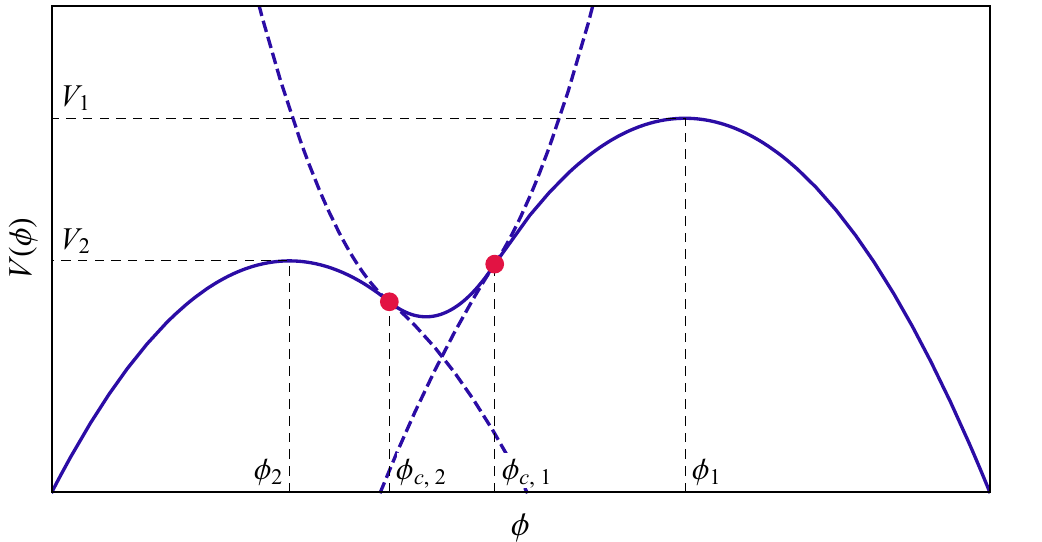}
    \caption{A sketch of the non-instantaneous SR to USR transition model potential \eqref{eq:V_humps_2}.}
    \label{fig:box_V_sketch}
\end{figure}

Similarly to the instantaneous transition case, we can build a potential that generates the behaviour~\eqref{eq:box}. Taking hints from section~\ref{sec:delta_potential}, it is clear that the potential should consist of three parabolas: one for the initial SR and another for the final CR phases as before, and one with a positive second derivative to connect them smoothly. We write it as
\be \label{eq:V_humps_2}
    V = \left\{
    \begin{array}{ll}
	 V_1\qty(1+\frac{1}{2}\eta_{V1}\qty(\phi-\phi_{1})^2) \, , &  \quad \phi_{c,1} < \phi \\
	 \tilde V\qty(1+\frac{1}{2}\tilde\eta_V\qty(\phi-\tilde\phi)^2) \, , &  \quad \phi_{c,2} < \phi \leq \phi_{c,1}  \\
	 V_2\qty(1+\frac{1}{2}\eta_{V2}\qty(\phi-\phi_{2})^2) \, , & \quad \phi \leq \phi_{c,2} 
    \end{array}	
    \right. \, , 
\ee
with $\lambda_i$ and $\eta_{Vi}$ related by Eq.~\eqref{eq:lambda_in_etaV}, and $\tilde{\lambda} = (3/2)\sqrt{(4/3)\tilde{\eta}_V - 1}$. Requiring $\calH^{-2}z''/z < 0$ in the intermediary region corresponds to $\tilde\eta_V > 2/3$. This makes sense: to break the SR approximation, a value of $\eta_V$ of order one is required. This time, $\eta_V$ has no delta peaks, only discontinuities, and thus the first derivative of the potential must be continuous everywhere. A sketch of the potential is depicted in Fig.~\ref{fig:box_V_sketch}.

Again, spectra of the form \eqref{eq:PR_box} can be produced by a potential of the form \eqref{eq:V_humps_2}. In the intermediary region, a solution similar to \eqref{eq:phi_humps_solution} applies, and the parameters of the $z$ and $V$ functions there can be straightforwardly matched by expressions similar to \eqref{eq:match_c_expressions}, paying attention to the continuity of the field velocity. In principle, even more complicated forms of $\calH^{-2}z''/z$ and the potential can be built and matched with a similar procedure. The details are not very illuminating, and we will not study them here.

The advantage of the ``box" model for non-instantaneous transitions is that it reduces the discontinuity of $\calH^{-2}z''/z$ and thus dampens the oscillations in the power spectrum peak. An example of this is shown in Fig.~\ref{fig:model_1_fits}, where the duration of the transition phase, $\Delta N_{\rm T} = N_2-N_1$, is varied while fitting the rising and falling slopes of the peak to the smooth Model I. The correct steepness of the slopes was achieved by setting $\lambda_1 = 1.8$ and $\lambda_2=1.916$. To fix the position of the falling slope, the end of USR with $z'=0$ was matched in the smooth model and in the non-instantaneous approximation. To fix the position of the rising slope, the position of the `dip' in $\calH^{-2}z''/z$ was varied by hand. The results show that quick transitions mimic the instantaneous approximation, as expected, while widening the transition smoothens the spectrum, but only to a degree. A spurious `bump' remains on top of the smooth spectrum, even for wide transitions, and the bump is further modulated by oscillations arising from the sharp edges at $N_1$ and $N_2$.

\section{Phenomenology}
\label{sec:pheno}

Aided by the analytical expressions for the curvature power spectrum, we will next estimate the abundance of PBHs and GWs induced by the large fluctuations, and consider the constraints from CMB.

\subsection{Primordial black holes}
\label{sec:PBHs}

High amplitude curvature perturbations lead to formation of PBHs after their horizon re-entry~\cite{Hawking:1971ei,Carr:1974nx,Carr:1975qj}. PBH formation occurs if the smoothed density contrast $\delta_m$ exceeds a threshold value $\delta_c$~\cite{Carr:1975qj,Musco:2004ak, Polnarev:2006aa, Musco:2008hv, Musco:2012au, Musco:2018rwt, Kehagias:2019eil, Musco:2020jjb}. We estimate the fraction of the total energy density $\beta_k(M)\td \ln M$ that collapses into BHs of mass $M$ using the Press-Schechter formalism~\cite{Press:1973iz,Bond:1990iw,Carr:1975qj,Gow:2020bzo}:
\be \label{eq:betak_def}
    \beta_k(M) 
    = \int_{\delta_{l,c}}^\infty \td\delta_l\, \frac{M}{M_k} \, P_k(\delta_l) \delta_D\!\left[ \ln\frac{M}{M(\delta_m(\delta_l))} \right] \,,
\ee
where $\delta_D$ denotes the Dirac delta function, $\delta_{l,c}$ is the threshold value of $\delta_l$, called the linear Gaussian component in the literature, which is related to the density contrast $\delta_m$ via~\cite{Young:2019yug,DeLuca:2019qsy,Kawasaki:2019mbl,DeLuca:2022rfz}
\be
\delta_{m} = \delta_l - \frac38 \delta_l^2 \,,
\ee 
and $P_k$ is its probability distribution. Using the critical scaling~\cite{Choptuik:1992jv,Niemeyer:1997mt,Niemeyer:1999ak}
\be\label{eq:Mcrit}
    M(\delta_m) = \kappa M_k \left(\delta_{m}  - \delta_c \right)^{\gamma} \,,
\ee
where $\gamma = 0.36$~\cite{Choptuik:1992jv,Evans:1994pj} and $M_k$ is the horizon mass
\be \label{eq:M_k}
    M_k = \frac{4\pi\mpl^2}{H_k} 
    \approx 1.4 \times 10^{13}\Msun \left(\frac{k}{\Mpc^{-1}} \right)^{-2}\, \left(\frac{g_{*,s}^{4}g_{*}^{-3}}{106.75}\right)^{-1/6} \,,
\ee 
we can express $\beta_k(M)$ as
\be \label{eq:betak}
    \beta_k(M) = \frac{2\kappa}{\gamma} \frac{q^{1+1/\gamma} P_k(\delta_l(M)) }{1 - \frac{3}{4}\delta_l(M)} \,,
\ee
where $q \equiv M/(\kappa M_k)$ and $\delta_l(M)$ is the inversion of the critical scaling law~\eqref{eq:Mcrit},
\be
    \delta_l(M) = \frac43\left[1 - \sqrt{1 - \frac{3}{2}\left(\delta_c + q^{1/\gamma}\right)}\right] \,.
\ee
The threshold density contrast $\delta_c$ and the $\kappa$ parameter depend on how the primordial perturbations are smoothed~\cite{Young:2019osy,Young:2020xmk,Gow:2020bzo} as well as on the shape of individual peaks~\cite{Musco:2018rwt,Young:2019yug,Escriva:2020tak,Escriva:2021pmf}. We use $\delta_c = 0.55$ and $\kappa=4.0$~\cite{Young:2019yug}. 

We assume that the curvature perturbations are Gaussian,
\be \label{eq:delta_prob_dist}
P_k(\delta_l) = \frac{1}{\sqrt{2\pi}\sigma_k} \,e^{-\delta_l^2/2\sigma_k^2} \,,
\ee
where $\sigma_k$  is the variance at scale $k$. The assumption of Gaussianity introduces potential inaccuracies as the curvature perturbations can develop non-Gaussianities in the form of an exponential tail in the probability distribution \eqref{eq:delta_prob_dist}, which can lead the Gaussian expression to underestimate the PBH abundance by several orders of magnitude~\cite{Pattison:2017mbe,Ezquiaga:2019ftu,Figueroa:2020jkf,Figueroa:2021zah}.
The correction is model-dependent and requires a non-linear analysis. There is currently no simple semi-analytical way to compute it accurately. We use \eqref{eq:delta_prob_dist} as a first approximation, keeping in mind that \eg~tuning the power spectrum height slightly can lead to comparable changes in the abundance. Neglecting the contribution of connected correlation functions of the curvature power spectrum~\cite{DeLuca:2022rfz}, the variance of $\delta_l$ is obtained from the curvature power spectrum $\PR(k)$~\cite{Young:2019yug}, 
\be \label{eq:sigmak}
    \sigma_k^2 = \left(\frac{4}{9}\right)^2\int_0^\infty \frac{\td k'}{k'}\left(\frac{k'}{k}\right)^{\!4} W^2(k'/k) T^2(k'/k) \PR(k') \,.
\ee
Following Ref.~\cite{Young:2019yug}, we evaluate the variance using a real-space top-hat window function 
$W(x) = 3 (\sin x - x \cos x)/x^3$, and accounting for the damping of sub-horizon fluctuations with the linear transfer function $T(x) = 3 [\sin(x/\sqrt{3}) - x/\sqrt{3} \cos(x/\sqrt{3})]/(x/\sqrt{3})^3$.
Finally, the present day PBH mass function, normalised to the PBH abundance, $\int \td \ln M \psi(M) = \Omega_{\rm PBH}$, is
\bea \label{eq:psiRD}
     \psi(M) 
&     = \int \td\ln k \, \beta_k(M) \frac{\rho_{\gamma}(T_k)}{\rho_c} \frac{s(T_0)}{s(T_k)} \, \\    
&     \simeq \frac{4 \times 10^{-12}}{\gamma} \!\frac{M}{\Msun} \int \!\frac{\td k \,k^2}{\Mpc^{-3}}  \frac{q^{1/\gamma}P_k(\delta_l(M))}{1 - \frac{3}{4}\delta_l(M)}  \,,
\eea
where $\rho_{\gamma}(T)$ and $s(T)$ are the radiation energy and entropy densities, and $\rho_c$ denotes the critical energy density of the Universe.

\begin{figure}[t]
    \centering
    \includegraphics[height=0.4\textwidth]{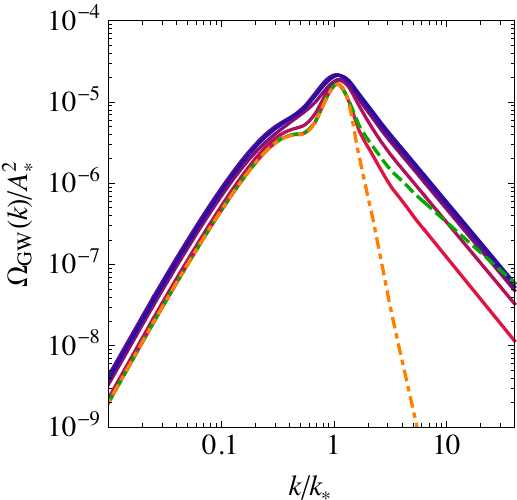}
    \includegraphics[height=0.4\textwidth]{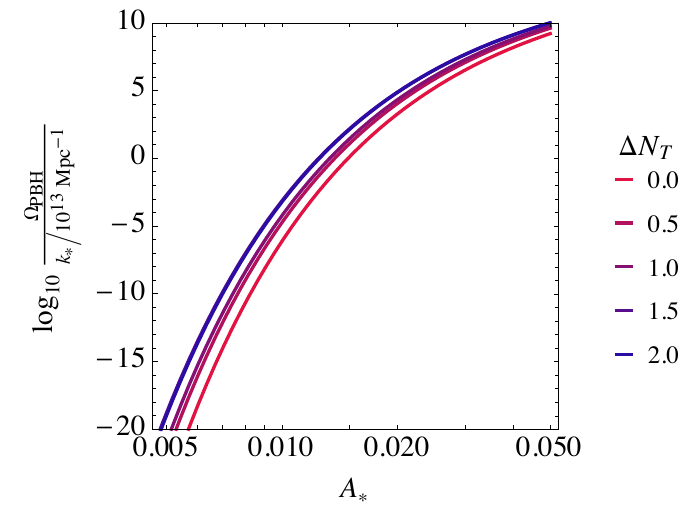}
    \caption{The abundance of gravitational waves (left) and primordial black holes (right). The solid curves correspond to $\lambda_2 = 1.916$ as in Fig.~\ref{fig:model_1_fits}, the green dashed to $\lambda_2 = 1.8$ and the orange dot-dashed to $\lambda_2 = 3.0$. The latter two correspond to the benchmark cases shown in Fig.~\ref{fig:Akplot}, and they overlap with the $\Delta N_T = 0$ curve in the right panel. For each case $\lambda_1 = 1.52$.}
    \label{fig:Omega}
\end{figure}

Similarly as in Ref.~\cite{Vaskonen:2020lbd}, we find that for the curvature power spectrum~\eqref{eq:PR_delta}, independently of $\lambda_2$, the PBH mass function is of the form
\be \label{eq:mf}
    \psi(M)\propto M^{1+1/\gamma} e^{-c_1 (M/\langle M_{\rm PBH} \rangle )^{c_2}} \,,
\ee
where $c_1$ is fixed such that $\langle M_{\rm PBH} \rangle$ is the average PBH mass, and $c_2 \simeq 1$ has a mild dependence on the peak amplitude of the curvature power spectrum. The abundance of PBHs and their mean mass are given in terms of the curvature power spectrum peak position $k_*$ and amplitude $A_*$ by
\bea \label{eq:fits}
    \Omega_{\rm PBH} &\simeq c_\Omega A_*^{c'_\Omega}\, e^{-c_A/A_*}\, k_*/\Mpc^{-1} \,,
\\ 
    \left \langle M_{\rm PBH} \right\rangle &\simeq c_M A_*^{c'_M} M_{k_*} \,,
\eea
where $c_\Omega \approx 30$, $c'_\Omega \approx 1.3$, $c_A \approx 0.4$, $c_M \approx 16$ and $c'_M \approx 0.34$. As shown in the right panel of Fig.~\ref{fig:Omega}, a non-instantaneous SR to USR transition only slightly changes the value of $A_*$ that gives rise to the particular PBH abundance.

In Fig.~\ref{fig:Akplot} we use the spectrum~\eqref{eq:PR_delta} for instantaneous SR to USR transitions and plot the projection of the PBH constraints in the plane of the amplitude $A_*$ and position $k_*$ of the maximum in the curvature power spectrum. The gray region in is excluded by the PBH constraints. These constraints arise from Hawking evaporation during big-bang nucleosynthesis and non-observation of extragalactic gamma-ray background from 
PBH evaporation~\cite{Carr:2009jm,Acharya:2020jbv}, 
microlensing results from Subaru/HSC~\cite{Niikura:2017zjd,Smyth:2019whb}, EROS~\cite{Tisserand:2006zx}, OGLE~\cite{Niikura:2019kqi}, Kepler~\cite{Griest:2013aaa} and MACHO~\cite{Allsman:2000kg},  
GW event rate observed by LIGO-Virgo~\cite{Hutsi:2020sol}, 
lensing of type Ia supernovae~\cite{Zumalacarregui:2017qqd}, 
lensing of LIGO-Virgo events~\cite{Urrutia:2021qak}, 
survival of stars in dwarf galaxies~\cite{Brandt:2016aco,Koushiappas:2017chw}, 
survival of wide binaries~\cite{Monroy-Rodriguez:2014ula}, 
Lyman-$\alpha$ forest data~\cite{Afshordi:2003zb,Murgia:2019duy} 
and limits on accretion~\cite{Ricotti:2007au,Horowitz:2016lib,Ali-Haimoud:2016mbv,Poulin:2017bwe,Hektor:2018qqw,Hutsi:2019hlw,Serpico:2020ehh}. The envelope of these constraints shown by the gray curve is calculated using the method introduced in~\cite{Carr:2017jsz} from the constraints for monochromatic mass functions to the mass function~\eqref{eq:mf}.

\subsection{Gravitational waves}
\label{sec:GWs}

\begin{figure}
    \centering
    \includegraphics[width=0.98\textwidth]{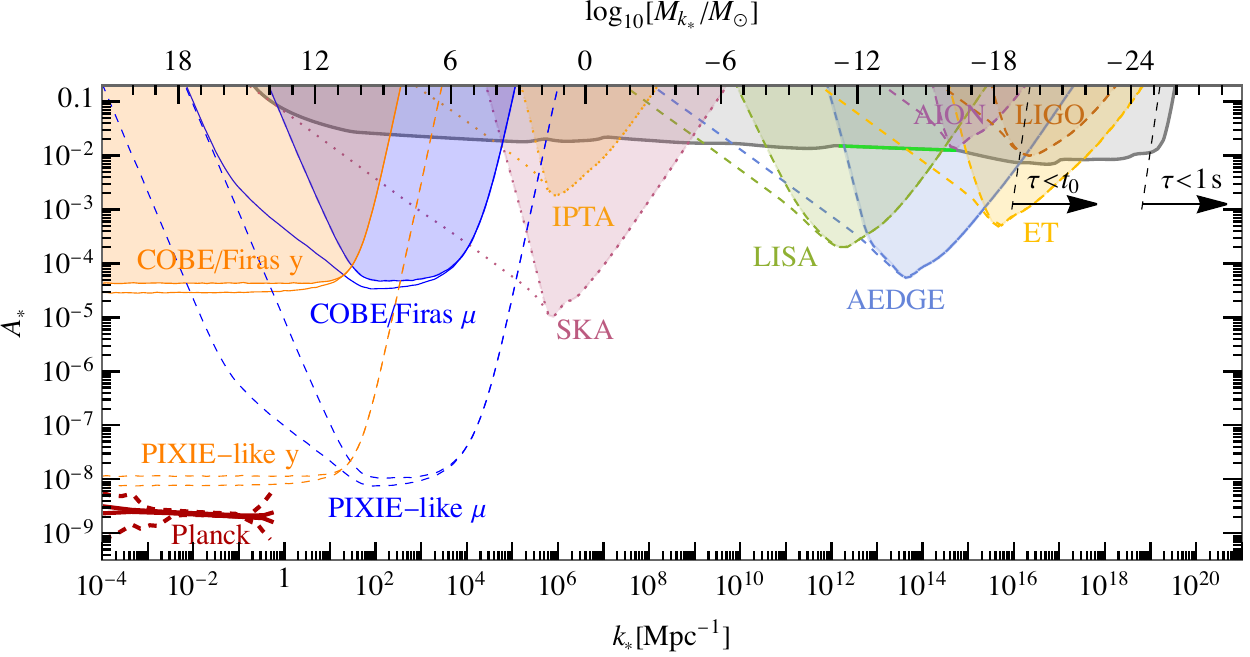}
    \caption{Constraints (solid lines) and projected sensitivities of future experiments (dashed lines) on the peak amplitude $A_*$ of the curvature power spectrum as a function of the peak position $k_*$ in the case of an instantaneous SR to USR transition. The red solid and dashed lines in the bottom left corner correspond to $1\sigma$ and $3\sigma$ constraints from Planck CMB observations. The gray solid line corresponds to the envelope of the PBH constraints and the green line highlights the asteroid mass window where all DM could be in PBHs. The PBH lifetime $\tau$ is evaluated for the average PBH mass. The parameters of the curvature power spectrum~\eqref{eq:PR_delta} are fixed to $\lambda_1=1.52$ and two cases for $\lambda_2$ are shown: $\lambda_2=1.8$ (smaller $A_*$ or weaker constraint) and $\lambda_2=3.0$ (larger $A_*$ or stronger constraint).}
    \label{fig:Akplot}
\end{figure}

Curvature perturbations induce formation of GWs at second order from the mode coupling~\cite{Matarrese:1993zf,Matarrese:1997ay,Nakamura:2004rm,Ananda:2006af,Baumann:2007zm}. Recently, these scalar-induced GWs (SIGWs) have been extensively studied (see for instance~\cite{Domenech:2021ztg,Kohri:2018awv,Inomata:2019yww,DeLuca:2019ufz,Yuan:2019fwv,Atal:2021jyo}), and the prospects for observing them have been considered~\cite{Saito:2008jc,Assadullahi:2009jc,Bugaev:2010bb,Alabidi:2012ex,Inomata:2016rbd,Orlofsky:2016vbd,Espinosa:2018eve,Inomata:2018epa,Byrnes:2018txb,Cai:2018dig,Bartolo:2018rku,Wang:2019kaf,Yuan:2019udt,Chen:2019xse,Lewicki:2021xku,LISACosmologyWorkingGroup:2022jok}. SIGWs have been searched for but not detected by the LIGO-Virgo collaboration~\cite{Kapadia:2020pnr,Romero-Rodriguez:2021aws}. However, it has been suggested that the recent NANOGrav result~\cite{NANOGrav:2020bcs} could be explained by SIGWs~\cite{Vaskonen:2020lbd,DeLuca:2020agl,Kohri:2020qqd,Domenech:2020ers}.

During radiation domination\footnote{We note that the SIGWs generally depend on the expansion history~\cite{Hajkarim:2019nbx,Inomata:2019ivs,Inomata:2019zqy,Domenech:2019quo,Domenech:2020kqm}.}, GWs decouple from scalar perturbations soon after horizon crossing and their abundance reaches a constant value. The SIGW spectrum today is given by (see e.g.~\cite{Kohri:2018awv,Inomata:2019yww})
\be
    \Omega_{\rm GW}(k) = 0.387\, \Omega_{\rm R} \left(\frac{106.75}{g_{*,s}^{4}g_{*}^{-3}}\right)^{\!\frac13}
    \frac{1}{6} \int_{-1}^1 \!\td x \int_1^\infty \!\td y \, \PR\!\left(\frac{x-y}{2}k\right) \PR\!\left(\frac{x+y}{2}k\right) F(x,y) \,,
\ee
where $\Omega_{\rm R} = 5.38\times10^{-5}$ is the radiation abundance~\cite{Planck:2018vyg}, the effective number of degrees of freedom are evaluated at the moment when the constant abundance is reached, roughly coinciding with the horizon crossing moment, and
\bea
    F(x,y) = &\frac{288(x^2+y^2-6)^2(x^2-1)^2(y^2-1)^2}{(x-y)^8(x+y)^8} \\ 
    &\times\left[\left(x^2-y^2+\frac{x^2+y^2-6}{2}\log\left|\frac{y^2-3}{x^2-3}\right|\right)^2 \!+ \frac{\pi^2(x^2+y^2-6)^2}{4}\theta(y-\sqrt{3}) \right] .
\eea
In the left panel of Fig.~\ref{fig:Omega} we show the GW spectrum corresponding to the curvature power spectrum~\eqref{eq:PR_delta} for three benchmark cases. In each case we have fixed $\lambda_1 = 1.52$ to match the CMB observations, and for $\lambda_2$ we consider three values, $1.8$ (green dashed), $1.916$ (solid) and $3.0$ (orange dot-dashed). Moreover, for $\lambda_2 = 1.916$, we consider the GW spectra and PBH abundances for the curvature power spectrum of a non-instantaneous SR to USR transition~\eqref{eq:PR_box} (compare to Fig.~\eqref{fig:model_1_fits}).

For a given shape of the peak in the curvature power spectrum, we can project the sensitivities of different GW observatories on the plane of the curvature power spectrum peak position $k_*$ and amplitude $A_*$. In Fig.~\ref{fig:Akplot} we show in two benchmark cases of an instantaneous USR to SR transition, corresponding to $\lambda_1 = 1.52$, $\lambda_2 = 1.8$ and $\lambda_1 = 1.52$ and $\lambda_2 = 3.0$, these projections for IPTA~\cite{Antoniadis:2022pcn}, SKA~\cite{Janssen:2014dka} LISA~\cite{LISA:2017pwj}, AEDGE~\cite{AEDGE:2019nxb,Badurina:2021rgt}, AION~\cite{Badurina:2019hst,Badurina:2021rgt}, ET~\cite{Sathyaprakash:2012jk} and LIGO~\cite{LIGOScientific:2014pky}.\footnote{In Fig.~\ref{fig:Akplot} we show the prospects for probing SIGWs with different future GW detectors. The prospects for probing the GWs from PBH binaries has been considered e.g. in Refs.~\cite{DeLuca:2021hde,Mukherjee:2021itf,Pujolas:2021yaw,LISACosmologyWorkingGroup:2022jok,Ng:2022agi,Franciolini:2022htd}.} We characterize the detectability of the GW signal by the signal-to-noise ratio
\be
    {\rm SNR} = \sqrt{\mathcal{T} \int \td f \left[\frac{\Omega_{\rm GW}(f)}{\Omega_{\rm sens}(f)}\right]^2} \,,
\ee
where $\mathcal{T}$ is the observation time and $\Omega_{\rm sens}(f)$ the sensitivity of the given detector. We take $\mathcal{T} = 5\,$years and ${\rm SNR}> 10$ for the threshold signal-to-noise ratio. By varying $\Delta N_{\rm T}$ in the toy model for non-instantaneous transitions~\eqref{eq:PR_box}, we checked that the sensitivities shown in Fig.~\ref{fig:Akplot} are only marginally affected by the duration of the transition.

The values $\lambda_2 = 1.8$ and $\lambda_2 = 3.0$ enclose most of the phenomenologically interesting range. For instance, Table~\ref{tab:delta_models} together with Eq.~\eqref{eq:lambda_in_etaV} gives the values $\lambda_2 = 1.83$ and $\lambda_2 = 2.05$---these correspond to spectra with relatively flat $\PR$ peaks producing solar mass PBHs and steeper peaks required for populating the asteroid mass PBH DM window, respectively. However, the identification $\lambda_2$ with the PBH mass is not unique and depends on other features of the inflationary scenario, such as the duration of inflation and whether it ends after the dual USR+CR phase during which the peak is produced (compare, for instance, Model I and Model III in Fig.~\ref{fig:example_models}).

\subsection{CMB constraints}
\label{sec:CMB}

The CMB observations directly strongly constrain the curvature power spectrum at scales $10^{-4}\,\Mpc^{-1} \lsim k \lsim 1\,\Mpc^{-1}$. In Fig.~\ref{fig:Akplot} we show with dark red curves the $1\sigma$ and $3\sigma$ Planck constraints on the evolution of the curvature power spectrum from Ref.~\cite{Planck:2018nkj}.

At redshifts $z\lesssim 10^6$ energy injections into the primordial plasma cause persisting spectral distortions in the CMB. These distortions are divided into chemical potential $\mu$-type distortions created at early times and Compton $y$-type distortions created at $z\lesssim 5\times 10^4$. For a given curvature power spectrum $\PR(k)$ the $X = \mu, y$ spectral distortions are~\cite{Chluba:2012we,Chluba:2013dna}
\be
X = \int_{k_{\rm min}}^\infty \frac{\td k}{k}\, \PR(k) W_X(k) \,,
\ee
where $k_{\rm min} = 1\,\Mpc^{-1}$ and the window functions can be approximated by
\be
W_\mu(k) = 2.2 \left[e^{-\frac{(\hat{k}/1360)^2}{1 + (\hat{k}/260)^{0.6} + \hat{k}/340}} - e^{-\left(\hat{k}/32\right)^2} \right]\,, \qquad
W_y(k) = 0.4 e^{-\left(\hat{k}/32\right)^2}\,,
\ee
with $\hat{k} = k/(1\,\Mpc^{-1})$. 

The COBE/Firas observations constrain the $\mu$ and $y$ distortions as $\mu \leq 9\times 10^{-5}$ and $y \leq 1.5 \times 10^{-5}$~\cite{Fixsen:1996nj} at $95\%$ CL. These imply strong constraints on the curvature power spectrum in the range $k \lesssim 10^{4} \Mpc^{-1}$~\cite{Chluba:2012we,Chluba:2013dna}. Moreover, a PIXIE-like detector could probe distortions as small as $\mu \leq 2\times 10^{-8}$ and $y \leq 0.4 \times 10^{-8}$~\cite{Kogut:2011xw}. We show the COBE/Firas constraints as well as the projections of a PIXIE-like detector sensitivity in Fig.~\ref{fig:Akplot} for two benchmark cases of the curvature power spectrum~\eqref{eq:PR_delta}. We find that the $\mu$-distortion constraints exclude single-field formation of PBHs heavier than $10^4 \Msun$. Importantly, we find that this constraint is negligibly affected by variations in model parameters.

\section{Conclusions}
\label{sec:concl}

We have considered the general structure of single-field inflationary models capable of producing peaked curvature spectra. We have restricted our study to models in which an Einstein frame exists and the dynamics can be captured by a single function---the Einstein frame potential. In such cases, both numerical and analytic approaches indicate that inflation can be divided into the following epochs: 
\begin{itemize}
    \item \emph{Slow-roll}. Typically the density fluctuations at the CMB scale are produced in this phase.
    \item \emph{Slow-roll to ultra-slow-roll transition}. This phase will involve violation of the slow-roll conditions or even a brief exit from an inflationary cosmology. The power spectrum begins to grow slightly before this phase. The details of this phase vary the most from model to model. For instance, it can be triggered by a small maximum in the potential or by rolling through its global minimum.
    \item \emph{Ultra-slow-roll}. The field starts to rapidly lose its velocity either by climbing up a hill in the potential or by rolling down a near plateau. The power spectrum typically reaches its peak during this phase.
    \item \emph{Constant-roll} (or also slow-roll). The velocity of the field begins to grow as the field begins to roll down the potential constantly (or slowly). If the potential is continuous, this phase is dual to the previous one and sets the declining slope of the power spectrum peak.
\end{itemize}
We have omitted a transitory period from ultra-slow-roll to constant-roll from the above list because it is entirely described by the duality of its neighbouring phases. All listed phases are present in single-field inflationary models for PBHs and can contribute to the enhanced curvature power spectrum. 

Assuming instantaneous slow-roll to ultra-slow-roll transitions, we have constructed an analytically solvable scenario that accounts for the inherent duality between the consecutive ultra-slow-roll and constant-roll phases. Such transitions can arise from piecewise quadratic potentials. The resulting curvature power spectrum can be consistent with CMB observations and can include a high peak that leads to a significant abundance of PBHs of any mass. The curvature power spectrum in this case oscillates at small scales below the peak. Relying on our analytic estimates, we have devised a simple analytic prescription in Eq.~\eqref{eq:PR_delta_approx} that allows to approximate peaked power spectra for general single-field inflationary models.  

We have studied how the shape of the peak depends on details of the slow-roll to ultra-slow-roll transition by constructing simplified solvable models in which the duration of this transition can be easily varied. The oscillations found in the case of instantaneous transition get smoother for slower transitions. These models can be considered as limiting cases of smooth physical potentials. In particular, we conclude that, in general single-field models for PBHs, the tails of the peak in the power spectrum can be determined from the potential alone via the second potential slow-roll parameter, while the exact shape of the peak (for instance the presence of oscillating features) depends on the details of how the initial slow-roll phase transitions into the subsequent ultra-slow-roll phase.

We have considered how the shape of the peak in the power spectrum affects the PBH abundance and the related scalar-induced GWs. A firm grasp of this relation is crucial for probing PBH phenomenology via GW observations. We have found that, although the details of the peak shape can affect the PBH abundance by a few orders of magnitude, the prospects for the detection of scalar-induced GWs as well the power spectrum amplitude required for abundant PBH production are quite insensitive to it. Next-generation GW detectors such as LISA and AEDGE are expected to probe the entire asteroid mass window for PBH dark matter. The COBE/Firas $\mu$-distortion constraints exclude the production of PBHs heavier than $10^4 \Msun$ in single-field inflation.

\acknowledgments

This work was supported by the Estonian Research Council grants PRG803, PRG1055, PSG761, and MOBTT5, the EU through the European Regional Development Fund CoE program TK133 ``The Dark Side of the Universe", the Spanish MINECO grants IJC2019-041533-I, FPA2017-88915-P and SEV-2016-0588, the Spanish MICINN (PID2020-115845GB-I00/AEI/10.13039/501100011033), and the grant 2017-SGR-1069 from the Generalitat de Catalunya. IFAE is partially funded by the CERCA program of the Generalitat de Catalunya.

\appendix

\section{Potentials and parameters for the example models}
\label{app:examples}

The inflationary models given as an example in Fig.~\ref{fig:example_models} of section~\ref{sec:theory} 
have actions of the form
\bea
	S = \int \td^4 x\, \sqrt{-g} \left( \frac{1}{2}\mpl^{2} \Omega(\sigma) R - \frac{1}{2}K(\sigma) (\partial \sigma)^{2} - V(\sigma) \right)\, ,
\eea
where $\sigma$ is the inflaton, $\Omega(\sigma)$ contains the non-minimal couplings, $K$ accounts for the possibility of a non-canonical kinetic term, $V$ is the potential, $V(v) = 0$ at the vacuum with $\sigma=v$, and $\mpl =  2.4 \times 10^{18} \GeV$ is the reduced Planck mass. In the following sections, we specify these functions for the different models. All parameter values are gathered in table~\ref{tab:example_model_parameters}, and the inflationary observables are presented in table~\ref{tab:example_model_observables}. To compute the results, we have performed the appropriate field redefinitions and conformal transformations \cite{Jarv:2016sow} to eliminate $\Omega$ and $K$ and bring the action back to the standard form \eqref{eq:S_standard}.

\subsection{Model I: non-minimal polynomial inflation} \label{sec:poly_model}

This model is defined by
\be
    V = \frac{1}{2}m^2\sigma^2 + \frac{1}{3}\mu\sigma^3 + \frac{1}{4}\lambda\sigma^4, \qquad
    \Omega = 1 + \xi \sigma^2/\mpl^2, \qquad
    K = 1 \, ,
\ee
inspired by~\cite{Kannike:2017bxn, Ballesteros:2017fsr, Ballesteros:2020qam}. The cubic term can be used to create a small maximum in the potential. Following~\cite{Kannike:2017bxn}, we reparametrize the model in terms of the extrema of the Einstein frame potential $V_E \equiv V/\Omega^2$ at $\sigma=v_i$, $i=1,2,3$, giving
\be
    m^2 = \frac{1}{2} \lambda \frac{v_{2} v_{3} (\xi  v_{2} v_{3}+3\mpl^2)\mpl^2}{\xi ^2 v_{2}^2 v_{3}^2-\xi  (v_{2}-v_{3})^2 \mpl^2 +3 \mpl^4} \, , \qquad
    \mu = -\frac{\lambda(v_{2}+v_{3})\mpl^4}{\xi ^2 v_{2}^2 v_{3}^2-\xi  (v_{2}-v_{3})^2\mpl^2+3\mpl^4} \, .
\ee
We use $v_1 \leq v_2 \leq v_3$, where, by construction, the trivial extremum lies at the origin, $v_1 = 0$. An exact inflection point corresponds to $v_2 =v_3$. The second potential SR parameter at the local maximum $v_2$ is
\be
    \eta_{V,c} 
    = - \frac{4 (v_{3} - v_{2}) (\xi  v_{2} (v_{2}+2 v_{3})+3\mpl^2)\mpl^4}{v_{2}^2 \left(\xi  (6 \xi +1) v_{2}^2+\mpl^2\right) \left(2 v_{3}\mpl^2 - v_{2} \left(\mpl^2-\xi  v_{3}^2\right)\right)} \, .
\ee
By Eq.'s \eqref{eq:PR_CR} and \eqref{eq:lambda_in_etaV}, it gives the high $k$ slope of the $\PR$ peak. The observables of Fig.~\ref{fig:example_models} are computed in the Einstein frame, with the canonical field $\phi$ solved numerically from
\begin{equation} \label{eq:field_transforamtion}
    \frac{\dd \phi}{\dd \sigma} = 
    \frac{1}{\Omega(\sigma)}\sqrt{\Omega(\sigma) + \frac{3\mpl^2}{2}\qty(\frac{\dd \Omega(\sigma)}{\dd \sigma})^2} \, .
\end{equation}

\begin{table}
    \centering
    \begin{tabular}{ccccc}
    \toprule
    Model & \multicolumn{4}{c}{Parameters} \\
    \midrule
    I & $\lambda = 2.10\times 10^{-5}$ & $\xi = 80.0118$ & $v_2 = 0.08\mpl$ & $v_3 = 0.0855714\mpl$  \\
    II & $V_0 = 1.89\times10^{-10}\mpl^4$ & $\alpha=1$ & $A=0.130383$ & $f_\sigma = 0.129576$ \\
    III & $V_0=7.03\times 10^{-11}\mpl^4$ & $A=1.17\times10^{-3}$ & $\phi_0=2.188\mpl$ & $\Delta=1.59\times10^{-2}\mpl$ \\
    \bottomrule
    \end{tabular}
    \caption{Parameter values for the models I, II, and III plotted in Fig.~\ref{fig:example_models}. No digits have been omitted: all numerical computations were performed with these exact values. See sections \ref{sec:poly_model}, \ref{sec:alpha_attr_model}, \ref{sec:bump_model} for the expressions for the potentials.}
    \label{tab:example_model_parameters}
\end{table}

\subsection{Model II: \texorpdfstring{$\alpha$}{α}-attractor inflation} 
\label{sec:alpha_attr_model}

This model, from \cite{Dalianis:2018frf}, is defined in the setup of superconformal $\alpha$-attractor models, with the inflaton sector given by
\be
    V = f^2\qty(\frac{\sigma}{\sqrt{3}\mpl}), \qquad
    \Omega = 1, \qquad
    K = \frac{2\alpha}{\qty(1-\frac{\sigma^2}{3\mpl^2})^2} \, .
\ee
The kinetic term is canonical for the redefined field
\begin{equation}
    \phi = \sqrt{6\alpha}\mpl\tanh^{-1}\frac{\sigma}{\sqrt{3}\mpl} \, ,
\end{equation}
and we pick a model with $f(x) = \lambda\qty(x + A\sin \frac{x}{f_\sigma})$ and $\alpha=1$, so that the canonical potential reads
\begin{equation} \label{eq:alpha_attr_V}
    V(\phi) = V_0\qty[\tanh\qty(\frac{\phi}{\sqrt{6}\mpl}) + A\sin\qty(f_\sigma^{-1}\tanh\qty(\frac{\phi}{\sqrt{6}\mpl}))]^2 \, , \qquad V_0 = \lambda^2 \, .
\end{equation}
The sinusoidal modulation produces an inflection point to the otherwise plateau-like potential. In the notation of \cite{Dalianis:2018frf}, this is Model II, and we use their parametrization \#1, reproduced in table~\ref{tab:example_model_parameters}.

\subsection{Model III: modified KKLT inflation}
\label{sec:bump_model}

This model, from \cite{Mishra:2019pzq}, is a modified version of the string theory based KKLT inflation, defined by
\be
    V = V_0\frac{\phi^2}{(\mpl/2)^2 + \phi^2}\qty(1+\epsilon(\phi)) \, , \qquad
    \Omega = 1 \, , \qquad
    K = 1 \, .
\ee
Gravity and the kinetic term are canonical from the get-go. The base plateau potential is modified by a small Gaussian bump,
\begin{equation}
    \epsilon(\phi) = A \,e^{-(\phi-\phi_0)^2/(2\Delta^2)} \, ,
\end{equation}
and dynamics around this bump enhance the power spectrum. We use one of the parameter sets given in \cite{Mishra:2019pzq}.

\begin{table}[t]
    \centering
    \begin{tabular}{ccccccccc}
    \toprule
    Model  
    & $n_s$ & $r$ & $N_\text{CMB}$ & $\phi_\text{CMB}/\mpl$ & $k_\text{peak}$ (Mpc$^{-1}$) & $\PR{}_\text{peak}$ & $M_\text{PBH}$ (g) \\
    \midrule
    I  & $0.947$ & $0.023$ & $56.9$ & $6.78$ & $1.25\times10^{14}$ & $0.035$ & $1.79 \times 10^{18}$ \\
    II & $0.945$ & $0.0075$ & $56.7$ & $5.85$ & $9.09\times10^{13}$ & $0.017$ & $3.37\times10^{18}$ \\
    III & $0.968$ & $0.0022$ & $60.0$ & $3.08$ & $2.18\times10^{14}$ & $0.0016$ & $5.88\times10^{17}$ \\
    \bottomrule
    \end{tabular}
    \caption{Inflationary observables for the models I, II, and III. For models II and III, these differ slightly from those reported in the original articles \cite{Dalianis:2018frf} and \cite{Mishra:2019pzq}, presumably due to the limited numerical accuracy of the input parameter values given there.}
    \label{tab:example_model_observables}
\end{table}

\section{Solution of the non-instantaneous transition model}
\label{app:box}

The ansatz~\eqref{eq:box} for $z$ splits the background evolution into three periods as in Eq.~\eqref{eq:z_box},
\be
	z = \left\{
\begin{array}{ll}
    \zeta_1 e^{(\lambda_1 -\frac{1}{2})(N - N_1)}, 
    &  \quad N < N_1 
    \\
	\tilde\zeta_{+} e^{(i\tilde\lambda -\frac{1}{2})(N - N_1)} + \tilde\zeta_{-} e^{(-i\tilde\lambda -\frac{1}{2})(N -N_1)}, 
    &  \quad N_1 < N < N_2
    \\
	\zeta_{2+} e^{(\lambda_2-\frac{1}{2})(N-N_2)} + \zeta_{2-} e^{(-\lambda_2-\frac{1}{2})(N-N_2)} ,
    & \quad N_2 < N
\end{array}	\right. \, ,
\ee
where the $\zeta_i$ coefficients are determined by continuity of $z$ and $z'$. Their explicit forms are $\zeta_{1\pm}/\zeta_1 = (1 \mp i \lambda_1/\tilde\lambda)/2$ and
\be
    \frac{\zeta_{2\pm}}{\zeta_1}
    = e^{-\Delta N_{\rm T}/2}\frac{1}{2}
    \left[
        \left(1\pm\frac{\lambda_1}{\lambda_2}\right)\cos(\tilde\lambda \Delta N_{\rm T})
    +   \left(\frac{\lambda_1}{\tilde\lambda}\mp\frac{\tilde\lambda}{\lambda_2}\right)\sin(\tilde\lambda \Delta N_{\rm T})
    \right] \, .
\ee
The coefficient $\zeta_{2+}$ enters in the curvature power spectrum. The instantaneous case \eqref{eq:insta} is obtained in the limit $\Delta N_{\rm T} \to 0$ with $\Delta N_{\rm T} \tilde \lambda^2 \equiv \mathcal{A}$ kept constant.

The corresponding general solution of the MS equation~\eqref{eq:MS} is
\be
	u_{k} 
	= 
\left\{\begin{array}{ll}	
	\frac{ \sqrt{\pi}}{\sqrt{\calH} 2\sin(\pi \lambda_1)} \left[ 
	c_{1+} J_{-\lambda_1} \left(\frac{k}{\calH}\right) + 
	c_{1-} J_{\lambda_1} \left(\frac{k}{\calH}\right) \right], 
	&  \quad N < N_1 \\
	\frac{ \sqrt{\pi}}{\sqrt{\calH} 2\sin(\pi i\tilde\lambda)} \left[ 
	\tilde c_{+} J_{-i\tilde\lambda} \left(\frac{k}{\calH}\right) + 
	\tilde c_{-} J_{i\tilde\lambda} \left(\frac{k}{\calH}\right) \right], 
	&  \quad N_1 < N < N_2 \\
	\frac{ \sqrt{\pi}}{\sqrt{\calH} 2\sin(\pi \lambda_2)} \left[ 
	c_{2+} J_{-\lambda_2} \left(\frac{k}{\calH}\right) + 
	c_{2-} J_{\lambda_2} \left(\frac{k}{\calH}\right) \right], 
	&  \quad N_2 < N
\end{array}\right. \, .
\ee
The solution must match the Bunch-Davies vacuum~\eqref{eq:BD_cond} at $N\rightarrow -\infty$. This gives the solution at $N_1$,
\be
    u_k = \frac{1}{\sqrt{\calH}}\frac{\sqrt{\pi}}{2} e^{i(\nu+\frac{1}{2})\frac{\pi}{2}}H^{(1)}_\nu\left(k/\calH\right) \, , \qquad N<N_1.
\ee
The coefficients $c_{i}$ are solved by demanding continuity of $u_k$ and its first derivative. The coefficients are
\bea
    \tilde c_{\pm} & = \left.\mp\frac{\pi}{2}e^{i\left(\lambda_1+\frac{1}{2}\right)\frac{\pi}{2}}
\left[
    J'_{\pm i\tilde\lambda}H^{(1)}_{\lambda_1}-J_{\pm i\tilde\lambda}{H^{(1)}_{\lambda_1}}'
\right]\right|_{N=N_1},\\
    c_{2\pm} & = \mp\left. \frac{\pi}{2\sin(\pi i\tilde\lambda)}
\bigg[
    \tilde c_{+}\left(
    J_{-i\tilde\lambda}J'_{\pm\lambda_2} - J'_{-i\tilde\lambda}J_{\pm\lambda_2}
    \right)
+
    \tilde c_{-}\left(
    J_{i\tilde\lambda}J'_{\pm\lambda_2} - J'_{i\tilde\lambda}J_{\pm\lambda_2}
    \right)
\bigg]\right|_{N=N_2},
\eea
where the prime denotes differentiation with respect to the number of $e$-folds $N$. The coefficient $c_{2+}$ enters the power spectrum. It can be expressed as
\bea\label{c3PlusModified}
    c_{2+} 
    = \frac{\pi^2 k^2 e^{i\left(\lambda_1+\frac{1}{2}\right)\frac{\pi}{2}}}{4\calH_1 \calH_2 \sin(i \pi \tilde\lambda)}
&   \Bigg[(B_1 B_2)J_{\lambda_2}(x_2) H^{(1)}_{\lambda_1}(x_1)
    \\
&   + \Big((B_2 C_1)J_{\lambda_2}(x_2) H^{(1)}_{\lambda_1-1}(x_1)
    + (C_2 B_1)J_{\lambda_2-1}(x_2) H^{(1)}_{\lambda_1}(x_1)\Big)
    \\
&   + (C_1 C_2)J_{\lambda_2+1}(x_2) H^{(1)}_{\lambda_1-1}(x_1) - (i\tilde\lambda\leftrightarrow -i\tilde\lambda )\Bigg],
\eea
where $x_i \equiv k/\calH_i\equiv k/\calH(N_i)$, and
\bea
&   B_1 = \frac{1}{2}\left(1-\frac{\lambda_1}{i\tilde\lambda}\right)J_{-i\tilde\lambda-1}(x_1)
        - \frac{1}{2}\left(1+\frac{\lambda_1}{i\tilde\lambda}\right)J_{-i\tilde\lambda+1}(x_1),\\
&   B_2 = \frac{1}{2}\left(1-\frac{\lambda_2}{i\tilde\lambda}\right)J_{i\tilde\lambda-1}(x_2)
        -\frac{1}{2} \left(1+\frac{\lambda_2}{i\tilde\lambda}\right)J_{i\tilde\lambda+1}(x_2),\\
&   C_1 = -J_{-i\tilde\lambda}(x_1), \qquad
    C_2 = J_{i\tilde\lambda}(x_2).\\
\eea
The first term in Eq.~\eqref{c3PlusModified} is responsible for the initial CMB part of the power spectrum. However, it also contributes to the peak in the power spectrum. Although its contribution is relatively minor, it is responsible for removing the oscillations from the peak when the duration of the transitional period is sufficiently long.

In the limit $\Delta N_{\rm T} \to 0$ with $\Delta N_{\rm T} \tilde \lambda^2 \equiv \mathcal{A}$ fixed, we can use $J_{i\tilde\lambda}(x) \to (x/2)^{i\tilde\lambda}/\Gamma(1+i\tilde\lambda)$ and we recover $c_{2+}$ in the instantaneous case \eqref{c2PlusDelta}.

\bibliography{PBHinflation}

\end{document}